\documentclass[preprint,12pt]{aastex}
\usepackage{graphicx}
\usepackage{psfrag}
\usepackage{amsmath}
\newcommand{\captionsize}{\footnotesize}

\newcommand{\dEk}{\delta_{\rm E}}
\newcommand{\Eeta}{{\rm E}_\eta}
\newcommand{\Ekappa}{{\rm E}_\kappa}
\newcommand{\Enu}{{\rm E}_\nu}

\newcommand{\bu}{{\bf u}}
\newcommand{\bB}{{\bf B}}
\newcommand{\be}{\hat{{\bf e}}}  

\newcommand{\dd}{{\rm d}}
\newcommand{\ee}{{\rm e}}  
\newcommand{\ii}{{\rm i}}  
\newcommand{\sech}{{\rm sech\,}}
\newcommand{\U}{U_0}  
\newcommand{\bx}{b_x}
\newcommand{\by}{b_y}
\newcommand{\bz}{b_z}
\newcommand{\ux}{u_x}
\newcommand{\uy}{u_y}
\newcommand{\uz}{u_z}
\newcommand{\hbx}{\hat{b}_x}
\newcommand{\hby}{\hat{b}_y}
\newcommand{\hbz}{\hat{b}_z}
\newcommand{\hux}{\hat{u}_x}
\newcommand{\huy}{\hat{u}_y}
\newcommand{\huz}{\hat{u}_z}
\newcommand{\fk}{f_0}  

\begin{document}

\title{THE SUN'S MERIDIONAL CIRCULATION AND INTERIOR MAGNETIC FIELD}
\author{\sc T.~S.~Wood$^1$, J.~O.~McCaslin$^2$, and P.~Garaud$^1$}
\affil{$^1$Department of Applied Mathematics and Statistics, Baskin School of Engineering, University of California Santa Cruz}
\affil{$^2$Department of Mechanical Engineering, College of Engineering and Applied Science, University of Colorado Boulder}
\email{tsw25@soe.ucsc.edu}

\begin{abstract}
To date, no self-consistent numerical simulation of the solar interior
has succeeded in reproducing the observed thinness of the solar tachocline,
and the persistence of uniform rotation beneath it. Although it is
known that the uniform rotation can be explained by the
presence of a global-scale confined magnetic field,
numerical simulations have thus far failed to produce any solution where
such a field remains confined against outward diffusion. We argue that the
problem lies in the choice of parameters for which these numerical
simulations have been performed. We construct a simple analytical
magneto-hydrodynamic model of the solar interior and identify several
distinct parameter regimes. For realistic solar parameter values, our
results are in broad agreement with the tachocline model of
\citeauthor{GoughMcIntyre98}.
In this regime, meridional flows driven at the base of the
convection zone are of sufficient amplitude to hold back the interior
magnetic field against diffusion. For the parameter values used in
existing numerical simulations, on the other hand, we find that meridional flows
are significantly weaker and, we argue, unable to confine the interior
field. We propose a method for selecting parameter values in future
numerical models.

\end{abstract}


\section{INTRODUCTION}
\label{sec:intro}

The differential rotation of the Sun's convective envelope terminates abruptly
at the interface with the underlying radiative zone.
The transition to uniform rotation takes place across a stably-stratified
shear layer known as the solar tachocline \citep{SpiegelZahn92},
whose thickness is at most a few percent of the Sun's radius
\citep{Thompson-etal96,Kosovichev-etal97,Schou-etal98,ElliottGough99,Charbonneaual99,
BasuAntia03}.

More than two decades after being first observed \citep{JCDSchou88,Brownal89}
a complete theory of the dynamics of the tachocline is still lacking.
In particular, the thinness of the tachocline seems to be at odds with
the known angular momentum transport properties of stratified rotating
fluids, including advection by meridional circulations
\citep{SpiegelZahn92,Elliott97}. Such circulations have a tendency to burrow into the
radiative zone, transporting the convection zone's angular momentum and
thereby thickening the tachocline.
In order for the tachocline to remain thin, some additional mechanism must be present
that transports angular momentum latitudinally, and in such a way as to enforce uniform
rotation. In addition, the
angular velocity
of the radiative zone
inferred from helioseismology
lies within the range of
angular velocities observed at the solar surface,
indicating
that the spin-down of the surface by
magnetic
solar-wind drag
\citep{Schatzman62}
has been communicated throughout the solar interior.
This requires significant \emph{vertical} transport of angular momentum.
Again, the transport must be such as to enforce uniform rotation.
We call such transport ``frictional'', meaning down-gradient in angular velocity.

Several authors have argued that a combination of anisotropic
turbulence and internal wave breaking can provide the required
latitudinal and vertical angular momentum transport
\citep[e.g.][]{KumarQuataert97,Zahn-etal97,CharbonnelTalon05}.
However, both of these processes are known to have \emph{anti}-frictional
properties \citep[e.g.,][]{McIntyre94,GoughMcIntyre98}.
On this basis, \citeauthor{GoughMcIntyre98} argued
that no non-magnetic model can explain the observed
thinness of the tachocline and the persistence of uniform rotation within
the radiative zone.
On the other hand, the presence of a global-scale primordial magnetic field
provides a natural explanation for the interior's uniform rotation
\citep[e.g.,][]{Ferraro37,Mestel53,MestelWeiss87}.
Such a field enforces uniform rotation through the elasticity of the field lines
imparted by magnetic tension.
The thinness of the tachocline can also be explained by the presence of
such a field, provided that the field lines are very nearly horizontal
within the tachocline, so that angular momentum is transported
from low to high latitudes by Maxwell stresses
\citep{RudigerKitchatinov97,GoughMcIntyre98,MacGregorCharbonneau99,WoodMcIntyre11}.
We call such a field ``confined'', meaning that the time-averaged
mean field resides below the base of the convection zone.
An unconfined field, on the other hand, would cause the convection
zone's differential rotation to propagate into the radiative zone
\citep{MacGregorCharbonneau99,Garaud02,BrunZahn06}.

To explain the confinement of the magnetic field, some process must be
invoked to counteract the outward diffusion of the field
throughout
the Sun's lifetime. Global-scale meridional flows that downwell from the convection zone into the
radiative zone offer a possible mechanism for achieving magnetic field confinement
--- these are the same meridional flows that would lead to tachocline thickening
in the
absence of an interior magnetic field.
This confinement mechanism was first suggested by \citet{GoughMcIntyre98}
and later studied numerically by \citet{GaraudGaraud08}.
The flows are generated by gyroscopic pumping in the convection zone,
a mechanism which has been widely studied in the context of the Earth's atmosphere,
and more recently in an astrophysical context \citep{McIntyre07,GaraudAA09,GaraudBodenheimer10}.
In brief, the same angular momentum transport by Reynolds stresses that gives rise to the
differential rotation of the convection zone also drives a meridional circulation through angular momentum conservation \citep[see Figure 1 of][]{GaraudBodenheimer10}.
Such meridional circulations are a robust consequence of anisotropic angular momentum
transport, as originally discussed by \citet{Kippenhahn63}.

Although the Sun's meridional flows can be observed directly at the solar surface,
and inferred in the near-surface layers from local helioseismology
\citep[e.g.,][]{Haber-etal02,ZhaoKosovichev04,Gizonal10},
their amplitude and structure deeper within the
solar interior is currently unknown.
Theoretical and numerical models of the large-scale dynamics of the solar interior are
required in order to establish whether the gyroscopically pumped circulation is strong enough, and penetrates deeply enough into the radiative zone, to confine the interior magnetic field and thereby explain the observed tachocline structure.
Several such numerical studies have been performed
\citep{Garaud02,BrunZahn06,GaraudGaraud08,Strugarek-etal11},
but none has successfully reproduced the field-confinement scenario
of \citeauthor{GoughMcIntyre98}.
However, since computing limitations prevent any numerical model from
reaching true solar parameter values, this failure may simply reflect
the fact that the numerical models are not in the parameter regime of
relevance to the solar tachocline.

\defcitealias{GaraudAA09}{GAA09}
An alternative, analytical approach was proposed by
\citet{GaraudBrummell08}, and further developed by
\citet[][hereafter \citetalias{GaraudAA09}]{GaraudAA09}.
These preliminary models examined the amplitude and depth of penetration of global-scale
meridional flows into the radiative zone
in the absence of an interior magnetic field.
In these non-magnetic models, the burrowing of the meridional flows
was halted only by the presence of viscosity.
In the steady state, the meridional mass flux was found to decay exponentially
with depth below the radiative--convective interface on 
a lengthscale $\sim R_\odot/\sigma$, where
$R_\odot$ is the solar radius and 
\begin{eqnarray}
\sigma = \frac{N}{\Omega_\odot}\sqrt{\frac{\nu}{\kappa}}\,.
\label{eq:sigma}
\end{eqnarray}
Here $\nu$ is the viscosity, $\kappa$ is the
thermal diffusivity, $N$ is the buoyancy frequency,
and $\Omega_\odot$ is the mean solar rotation rate. 
This result shows how the burrowing tendency of the meridional flows
is strengthened by rotation, and weakened by stable
stratification and viscosity. 
Within the solar tachocline
$\sigma < 1$, and so viscosity alone
cannot prevent the meridional flows
from burrowing
into the radiative zone.

\defcitealias{GaraudBodenheimer10}{GB10}
\citet{GaraudBrummell08} found that
the amplitude of the steady-state meridional flows
in their model was sensitive not only to the value of $\sigma$, but also
to conditions at the interface between the convective
and radiative zones \citep[see also][]{BrethertonSpiegel68}.
In a subsequent study,
\citet{GaraudBodenheimer10}
showed that the
vertical mass flux below the radiative--convective interface
can be quantified in terms of two constraints.
The first constraint is that there must be
stresses present in the radiative zone that overcome Taylor--Proudman balance. 
In the absence of such stresses, any meridional flows would have
to be parallel to the rotation axis, which is not compatible
with mass conservation. When such stresses are present, however,
downwelling meridional flows are able to turn around and
return to the convection zone --- this is another example of gyroscopic pumping.
\citeauthor{GaraudBodenheimer10} called this the ``mechanical'' constraint.\footnote{
Since the mechanical constraint arises from the need to balance azimuthal
forces, it applies only to the steady state.
Transient meridional flows, such as those studied by \citeauthor{SpiegelZahn92},
are not subject to this constraint
and can be much stronger.
}

The second constraint is
due to the presence of stable stratification, which inhibits
vertical flows; we call this the ``thermal'' constraint.
\citet[][hereafter \citetalias{GaraudBodenheimer10}]{GaraudBodenheimer10}
found that this places an upper bound on
the vertical flow velocity $W$ of the form
\begin{equation}
|W| \le  \frac{R_\odot}{G \, t_{\rm ES}}
  \label{eq:thermal}
\end{equation}
where $t_{\rm ES}$
is the local Eddington--Sweet timescale,
\begin{equation}
	t_{\rm ES} = \left(\frac{N^2}{\Omega_\odot^2} \frac{R_\odot^2}{\kappa}\right)
	\label{eq:ES}
\end{equation}
\citep[e.g.,][]{SpiegelZahn92},
and $G$ is a dimensionless ``geometrical'' factor that depends on the details of the star's
internal structure.

The two constraints set two upper bounds for
the meridional flow velocity, which
then
scales as the smaller of the two.
In the case where the mechanical constraint is smaller,
the meridional flow velocity depends on the nature of the stresses that
overcome Taylor--Proudman balance.
\citetalias{GaraudBodenheimer10}
considered both laminar viscous stresses
and turbulent Reynolds stresses
in the convective cores of ``lithium-dip'' stars.
In the present paper we consider magnetic stresses, which we argue
are the dominant mechanism for angular momentum transport in
the solar radiative zone.
For simplicity, we neglect any turbulent stresses outside of
the convection zone.

Following \citetalias{GaraudAA09} and \citetalias{GaraudBodenheimer10}
we use a linearized, steady-state, Cartesian model of the solar interior
to study the interaction between meridional flows and
an imposed
confined magnetic field.
In \S\ref{sec:model} we describe the Cartesian model
and derive the governing equations.
In \S\ref{sec:unstrat} we present numerical and analytical solutions
in the case where stratification is absent.
We demonstrate a simple relationship between the structure and magnitude of the
confined magnetic field and the pattern and amplitude of the meridional flows.
In particular, we quantify the mechanical constraint from the confined magnetic field.
In \S\ref{sec:strat} we extend these results to the stratified case,
recovering the thermal constraint mentioned above.
We then discuss in \S\ref{sec:disc}
what insight the results of our simplified model
provide into the nonlinear dynamics of the solar interior.
In particular, we identify the various parameter regimes that occur in our model,
and we discuss in which of these regimes the predicted meridional flows
could plausibly bring about confinement of the magnetic field.
The results show that great caution must be exerted when interpreting the 
results of numerical simulations that use non-solar values for the governing parameters,
and provide guidance as to the design of future numerical simulations of the solar interior.

\section{A Cartesian model}
\label{sec:model}

We use a similar model to that of \citetalias{GaraudAA09}.
Although our model is intended as a simplified Cartesian analogue of the
tachocline picture proposed by \citet{GoughMcIntyre98}, it also
goes beyond theirs in certain respects. We explicitly model the thermal and
dynamical coupling between the convective and radiative zones, and are
thus able to draw more quantitative conclusions concerning
the role of this 
interaction in the global-scale solution. We also consider much wider
numerical ranges in the governing parameters,
and describe
the structure of the solutions in
several distinct
parameter regimes.
Our aim
is to identify in which parameter regime, if any,
the propagation of the differential rotation below the convection zone is
restricted
to a thin layer with a tachocline-like structure.

\subsection{Model geometry}

\begin{figure}[htbp]
\centering
\psfrag{y/pi}[][]{\footnotesize $y/\pi$}
\psfrag{z}{\footnotesize $z$}
\psfrag{U0}[]{\footnotesize $\U$}
\psfrag{B0}[][t]{\footnotesize $\bB_0(z)$}
\psfrag{Omega}{\footnotesize $\Omega_\odot$}
\psfrag{delta}[][]{\footnotesize $\delta=0.2$}
\psfrag{xlabel}[t][]{\footnotesize $B_0/{\cal B}_0=\ee^{-z/\delta}$}
\epsscale{0.8}
\plotone{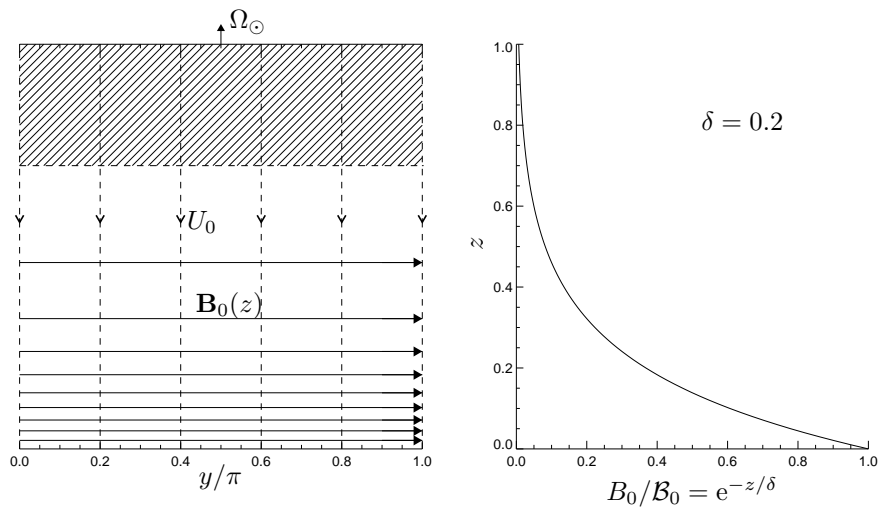}
\caption{\captionsize
  A meridional cross-section through the Cartesian model.
  The $y$ and $z$ axes represent latitude and altitude respectively.
  The shaded area marks the convection 
  zone, where a forcing is applied to mimic the generation of differential 
  rotation by turbulent stresses.
  The rotation axis is vertical.
  A global-scale
  magnetic field $\bB_0(z) = {\cal B}_0 \ee^{-z/\delta}\be_y$
  is confined by uniform downwelling $\U$.
  }
  \label{fig:schem}
\end{figure}

We measure distances in units of the solar radius, $R_\odot \simeq 7 \times 10^{10}$cm,
and velocities in units of $R_\odot\Omega_\odot$, where
$\Omega_\odot \simeq 3 \times 10^{-6}$\,s$^{-1}$ is the mean solar rotation rate,
which we use as the inverse timescale.
As illustrated in figure \ref{fig:schem},
the Cartesian coordinate system $(x,y,z)$
is chosen such that $x$ is the azimuthal coordinate,
$y\in[0,\pi]$ is the latitudinal coordinate, and $z\in[0,1]$ is the vertical coordinate.
Although there is no precise equivalence between our Cartesian
geometry and the spherical geometry of the solar interior,
for the sake of interpretation we will take
$y=0$ and $y=\pi$ to be the ``poles'', and $y=\pi/2$ to be the ``equator''.
The radiative--convective interface is at $z=h\simeq0.7$, so that
$h<z<1$ represents the convection zone, and $0<z<h$ represents the
radiative zone.

\subsection{Background state}

The steady background state considered is depicted in figure
\ref{fig:schem}.
We model the transition between the convective and radiative zones by imposing
a vertical profile for the dimensionless buoyancy frequency $n(z)$ with
\begin{equation}
n^2(z)=\frac{n^2_{\rm rz}}{2}\left\{1+\tanh\left(\frac{h-z}{\Delta}\right)\right\}\,,
\label{eq:bdef}
\end{equation}
so that $n=0$ in the convection zone and $n = n_{\rm rz} > 0$ in the radiative zone.
The lengthscale $\Delta$ is introduced in order to ensure smoothness of
the background state, which is necessary for numerical reasons. We may regard $\Delta$
as the thickness of the convective overshoot region at the base of the convection zone.

We impose a horizontal background magnetic field $\bB_0$ of the form
\begin{equation}
  \bB_0 \;=\; B_0(z)\be_y,
\end{equation}
where $\be_y$ is the unit vector in the $y$ direction. In order to confine this
field against ohmic diffusion, in the steady state, we must invoke a
background meridional flow.
The simplest case is that of uniform downwelling of dimensionless magnitude $\U$, 
and
\begin{equation}
B_0(z) \;=\; {\cal B}_0 \ee^{-z/\delta},
\end{equation}
where ${\cal B}_0$ is a constant and $\delta$ is the
dimensionless lengthscale
\begin{equation}
  \delta \;=\; \frac{\eta/(R_\odot^2\Omega_\odot)}{\U} \;=\; \frac{\Eeta}{\U}.
  \label{eq:U0def}
\end{equation}
Here $\eta$ and $\Eeta$ represent the magnetic diffusivity in
dimensional and nondimensional units respectively;
we call $\Eeta$ the ``magnetic Ekman number''.
Although this background state cannot hold over all latitudes and depths within
the solar interior (for reasons of mass conservation
and solenoidality of $\bB_0$),
we hope that it offers a reasonable qualitative approximation to conditions within
the tachocline
at middle and high latitudes.

We note that $\bB_0$ is not a force-free field since it has a non-constant
magnetic pressure $B_0^2/(8\pi)$.  However, the magnetic pressure can be
balanced by a perturbation to the background gas pressure $p_0$.
Estimates of the strength of the Sun's interior magnetic field
typically lie within the range $10^{-3}$--$10^{3}$ gauss
\citep[e.g.][see also \S\ref{sec:interpretation}]{MestelWeiss87,GoughMcIntyre98}.
For field strengths in this range, the gas pressure far exceeds the magnetic pressure,
and so the required perturbation is very small.

In what follows we consider linear, steady-state
perturbations to this background state.
The principle shortcoming of our model is, arguably,
the artificial manner in
which the magnetic field is confined to the interior, via the imposed
downwelling $\U$.
Magnetic confinement is thereby built into the model,
and controlled by the parameters ${\cal B}_0$ and $\delta$.
This idealization is
necessary to allow a linear
study.  A complete model of the tachocline would need to describe
self-consistently the processes that act to confine the magnetic field,
rather than treating it as part of the background.  Such a model would
necessarily be nonlinear, and awaits future numerical work.
We note also that our Cartesian model cannot adequately describe the
polar regions, where the effects of geometrical curvature become significant.
The polar magnetic confinement problem has been studied in detail elsewhere
\citep{WoodMcIntyre11}, and self-consistent, fully nonlinear solutions
have been obtained.

\subsection{Model equations}

As in the studies of \citetalias{GaraudAA09} and \citetalias{GaraudBodenheimer10} we do not model the turbulence in the convection zone in detail. Instead, we introduce a simple parametric model
to describe its role in driving a large-scale differential rotation profile,
and in enhancing the transport of heat. In the region $z \in [h,1]$
we model the turbulent heat transport as a diffusive process, and
we replace the divergence of the Reynolds stress
with a forcing term that represents linear relaxation towards
a prescribed rotation profile, which we choose to mimic
the observed differential rotation of the solar convection zone.

For simplicity we also make the Boussinesq approximation
\citep[e.g.][]{SpiegelVeronis1960},
in which
we neglect
variations in the background pressure $p_0$, density $\rho_0$ and temperature $T_0$,
and
we neglect
pressure perturbations
in the equation of state.
Although the Boussinesq approximation cannot be rigorously justified
over the entirety of our domain, we do not expect this approximation to have a
qualitative effect on our results. Indeed, we are interested primarily in solutions 
for which the meridional flows and differential rotation are confined to a thin, tachocline-like layer below
the convection zone,
within which the Boussinesq approximation should be quite accurate.

The nondimensional governing equations, 
linearized about the background state described in the previous section, are
\begin{align}
 2\be_z\times\bu &= -\nabla p + T \be_z
  + \Lambda\ee^{-z/\delta}\left[
    \frac{1}{\delta}\be_x \times \bB + (\nabla\times\bB) \times \be_y
  \right]  + \Enu\nabla^2\bu - \lambda(z)(\bu -\bu_{\rm cz}) \label{fullmom} \\
 n^2(z)\, w &= \Ekappa\nabla\cdot(f(z)\,\nabla T) \label{fullenergy} \\
 0 &= \nabla\times (
    \bu \times \ee^{-z/\delta}\be_y  - \frac{1}{\delta}\be_z \times\bB - \nabla\times\bB
  ) \label{fullind} \\
  \nabla\cdot\bu &= 0 \label{cont} \\
  \nabla\cdot\bB &= 0\,, \label{Bcont}
\end{align}
where $\bu = (\ux,\uy,\uz)$ is the velocity perturbation, $\bB = (\bx,\by,\bz)$
is the perturbation to the magnetic field, $p$ is the pressure perturbation,
and $T$ is the temperature perturbation.
These equations have been scaled
using $\rho_0R_\odot^2\Omega_\odot^2$ as the
unit of pressure, $T_0R_\odot \Omega_\odot^2/g$ as the unit of temperature,
and ${\cal B}_0/\Eeta$ as the unit of magnetic field strength.
The following dimensionless parameters have been introduced: 
\begin{eqnarray}
  \Lambda =&  \dfrac{{\cal B}_0^2 }{4\pi\eta\rho_0\Omega_\odot}\,,&
  \;\; \mbox{the Elsasser number at $z=0$;} \\
  \Enu =& \dfrac{\nu}{R_\odot^2\Omega_\odot}\,,&
  \;\; \mbox{the Ekman number;} \\
  \Ekappa =& \dfrac{\kappa}{R_\odot^2\Omega_\odot}\,,&
  \;\; \mbox{the thermal Ekman number in the radiative interior.}
\end{eqnarray}
Note that the non-dimensionalized perturbation equations
have no explicit dependence on the magnetic Ekman number $\Eeta$.
However, because $\Eeta$ has been used
to scale
the magnetic field, its value governs the magnitude of
perturbations relative to the background field $\bB_0$.

The Elsasser number $\Lambda$
is a measure of
the relative magnitude of Lorentz
forces compared to Coriolis forces.
It is convenient to define also a local
Elsasser number,
\begin{eqnarray}
  \Lambda_{\rm loc}(z) &=& \frac{B_0^2(z)}{4\pi\eta\rho_0\Omega_\odot} \\
  &=& \Lambda \ee^{-2z/\delta}\,,
\label{eq:Lambdaeff}
\end{eqnarray}
in order to measure the relative magnitudes of these forces at each altitude $z$.

In both the momentum equation (\ref{fullmom}) and the thermal energy
equation (\ref{fullenergy}) we have neglected terms describing advection
by the background downwelling $\U$.  It can be shown that, in the results
presented here, including these terms would produce only a small
correction to the solutions.  Neglecting these terms simplifies the analysis
and reduces the dependence of the solutions on the artificially imposed
flow $\U$.

The final term in the linearized momentum equation (\ref{fullmom}) describes a linear
relaxation towards a prescribed rotation profile $\bu_{\rm cz}$.  The
dimensionless parameter $\lambda$,
which determines the rate of relaxation,
can be regarded as the inverse of the convective turnover timescale.
This parameterization for the angular momentum transport by Reynolds stresses
is similar to that used by \citet{BrethertonSpiegel68}, \citetalias{GaraudAA09} and \citetalias{GaraudBodenheimer10}.
For simplicity, we assume that $\lambda$ takes the constant value $1/\tau_c$ in the
convection zone, and that the
prescribed
differential rotation $\bu_{\rm cz}$ is sinusoidal in latitude.
Following \citetalias{GaraudAA09}, we adopt the profiles
\begin{flalign}
&&\bu_{\rm cz}  &= u_{\rm cz}(z)\, \ee^{\ii ky}\be_x\,,& \label{ucz} \\
&\mbox{and}&  \lambda(z) &= \frac{1}{2\,\tau_c}\left\{1+\tanh\left(\frac{z-h}{\Delta}\right)\right\}\,,&
\label{eq:lambda}
\end{flalign}
where $\Delta$ is the same ``overshoot'' length as in equation (\ref{eq:bdef}).
In the numerical results presented later we set
$\tau_c = 0.1$ and $k=2$, representing equatorial symmetry
at the solar surface, and $u_{\rm cz}(z)$ is taken to be linear in $z$. We show in
appendix~\ref{sec:app-global} that the flow below the convection
zone is not very sensitive to the value of $\tau_c$, or to the form of $u_{\rm cz}(z)$,
and that the results are easily extended to any particular $u_{\rm cz}(z)$.
This insensitivity to the choice of parameterization within the convection zone
reflects the robustness of both the gyroscopic pumping mechanism and the burrowing
tendency of meridional flows.

Finally, we model turbulent heat transport within the convection zone
as an increase in the thermal diffusivity by a factor $f$
relative to its laminar, microscopic value.  We take
\begin{equation}
f(z) = 1 + \frac{\fk-1}{2}\left\{1+\tanh\left(\frac{z-h}{\Delta}\right)\right\}\,,
\label{D}
\end{equation}
so that $f=1$ within the radiative zone and $f=\fk \gg 1$ within the
convection zone. Thermal perturbations at the radiative--convective interface are therefore
communicated more efficiently into the convection zone than into the radiative
zone. In the limit $\fk \to \infty$ we expect the convection zone to become isothermal,
which is equivalent to isentropic in our Boussinesq model.
In the numerical results presented in \S\ref{sec:strat} we take $\fk=10^6$.
The physical significance of $\fk$, and its effect on the solutions,
is described in \S\ref{sec:interpretation}, below equation (\ref{eq:G1_G2}).

We assume that the viscosity $\nu$ and the magnetic diffusivity $\eta$ retain their laminar,
microscopic values throughout the solar interior.
Introducing larger ``turbulent'' values for $\nu$ and $\eta$ within the convection
zone would not significantly affect the results we present here.

The perturbation fields must have the same sinusoidal dependence on $y$ as the forcing
term $\lambda(z)\,\bu_{\rm cz}$ in (\ref{fullmom}),
and so we seek solutions of the form
\begin{equation}
q(y,z)=\hat{q}(z)\,\ee^{iky}\,.
\label{modes}
\end{equation}
The perturbation equations (\ref{fullmom})--(\ref{Bcont})
then become a system of ordinary differential equations
for the perturbation amplitudes $\hat{q}$:
\begin{eqnarray}
-2\huy &=& \ii k \Lambda \hbx\ee^{-z/\delta}+\Enu\left(\frac{\dd^2\hux}{\dd z^2}-k^2\hux \right) - \lambda(z)(\hux -u_{\rm cz}) \label{xmom} \\
2\hux &=& -\ii k\hat{p} - \frac{\Lambda}{\delta}\hbz\ee^{-z/\delta} + \Enu\left(\frac{\dd^2\huy }{\dd z^2}-k^2\huy \right) - \lambda(z)\huy \label{ymom} \\
\frac{\dd\hat{p}}{\dd z} - \hat{T} &=& \frac{\Lambda}{\delta}\hby\ee^{-z/\delta} + \Lambda\left(\ii k\hbz -\frac{\dd\hby }{\dd z}\right)\ee^{-z/\delta}+\Enu\left(\frac{\dd^2\huz }{\dd z^2}-k^2\huz \right) - \lambda(z)\huz \label{zmom} \\
0 &=& \ii k\hux \ee^{-z/\delta} + \frac{1}{\delta} \frac{\dd\hbx}{\dd z} + \frac{\dd^2\hbx}{\dd z^2}-k^2\hbx \label{xind} \\
0 &=& \ii k\huz \ee^{-z/\delta} + \frac{1}{\delta} \frac{\dd\hbz}{\dd z} + \frac{\dd^2\hbz}{\dd z^2}-k^2\hbz \label{intind} \\
\frac{n^2(z)}{\Ekappa}\huz  &=& \frac{\dd}{\dd z}\left(f(z)\frac{\dd\hat{T}}{\dd z}\right) - f(z)  k^2\hat{T} \label{ener} \\
\ii k\huy + \frac{\dd\huz}{\dd z} &=& 0 \label{divu} \\
\ii k\hby + \frac{\dd\hbz}{\dd z} &=& 0 \label{divb}\,.
\end{eqnarray}
The system of equations (\ref{xmom})--(\ref{divb}) is solved numerically
using a two-point boundary-value solver, given the boundary conditions listed below.

\subsection{Boundary conditions}
\label{sec:bcs}

The upper and lower boundaries are modeled as impenetrable, (viscous) stress-free,
isothermal and electrically insulating. That is, we impose
\begin{flalign}
  &&\left.
  \begin{array}{rcl}
    \huz &=& 0 \\
    \dfrac{\dd\hux}{\dd z} &=& 0 \\
    \dfrac{\dd\huy}{\dd z} &=& 0 \\
    \hat{T} &=& 0 \\
    \hbx &=& 0
  \end{array}
  \right\} \;\;\; &\mbox{at} \;\;\; z = 0 \;\; \mbox{and} \;\; z = 1,& \\
  &&\frac{\dd\hbz}{\dd z} = k\hbz \;\;\; &\mbox{at} \;\;\; z = 0,& \\
 &\mbox{and} &\frac{\dd\hbz}{\dd z} = -k\hbz \;\;\; &\mbox{at} \;\;\; z = 1.&
\end{flalign}
Most of these boundary conditions have been selected for 
simplicity and convenience.
In fact, the solutions turn out to be rather insensitive to the choice of
boundary conditions in the parameter regime of interest, as we have verified
by comparing solutions obtained with different sets of boundary conditions.

\section{The Unstratified Case}
\label{sec:unstrat}

\subsection{Numerical exploration}
\label{sec:numexp_unstrat}

In order to delineate the various physical effects present in the solutions,
we begin by considering the unstratified case, $n_{\rm rz}=0$.
Effects associated with stratification will be described in detail in \S\ref{sec:strat}.

\begin{figure}
\epsscale{0.7}
\plotone{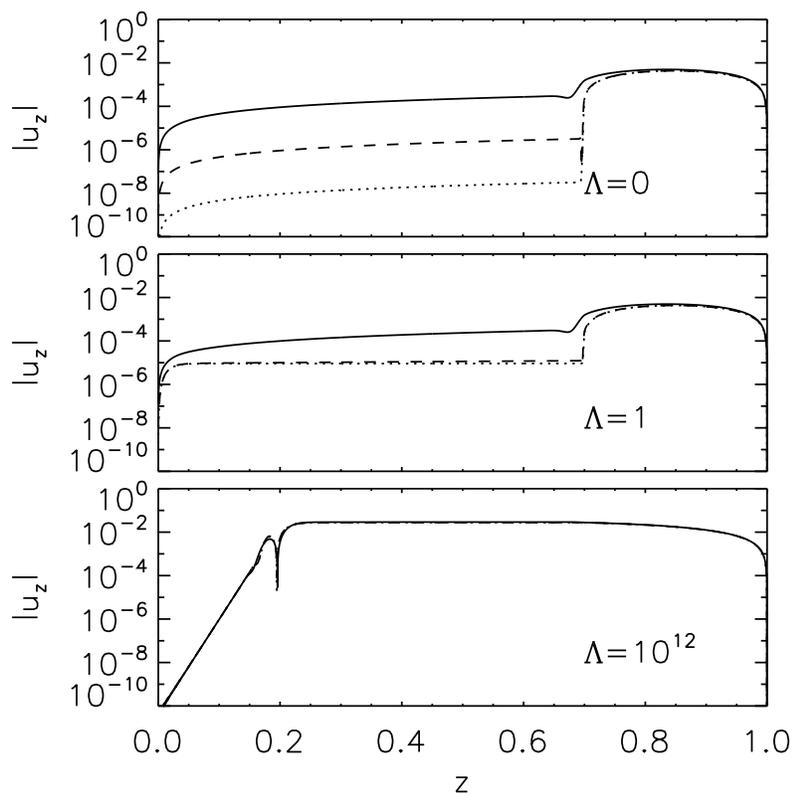}
  \caption{\captionsize Profiles of vertical velocity $\huz$
  for a series of unstratified solutions ($n_{\rm rz} = 0$)
  with different magnetic field strengths and Ekman numbers.
  All the solutions have $u_{\rm cz}(z) = 1+(z-h)$
  in (\ref{ucz}), and $\Delta = 0.001$ and $\tau_c = 0.1$ in
  (\ref{eq:lambda}).
  The magnetic scaleheight, as defined by (\ref{eq:U0def}), is $\delta = 0.02$. 
  Each panel shows the solutions for $\Enu = 10^{-4}$ (solid line),
  $\Enu=10^{-6}$ (dashed line) and $\Enu = 10^{-8}$ (dotted line).
  }
\label{E+Lambdavary}
\end{figure}

A series of results is shown in figure \ref{E+Lambdavary}.
In the absence of any magnetic field (i.e., setting $\Lambda = 0$) we  recover the non-magnetic, unstratified solution of \citetalias{GaraudAA09} (see their figure~2).
Since the lower boundary is stress-free, the differential rotation imposed within the convection
zone is able to spread throughout the radiative zone,
and establishes a state close to Taylor--Proudman balance.
In this steady state the only meridional flows are those maintained by viscosity,
and their magnitude scales with the Ekman number, $\Enu$.
The meridional flows therefore become vanishingly weak
in the limit $\Enu \to 0$. When $\Lambda >0$, on the other hand,
the Lorentz force produces departures from Taylor--Proudman balance,
and maintains a meridional flow even in the limit $\Enu \to 0$
(figure \ref{E+Lambdavary}b).

For $\Lambda$ of order unity, the Lorentz forces are most significant close to
the lower boundary $z=0$, within a boundary layer analogous to an
Ekman--Hartmann layer.
The structure of the boundary layer depends on the particular choice of
boundary conditions at $z=0$, including the magnetic boundary conditions.
However, we have verified that the meridional flow outside the boundary layer
is hardly affected by the choice of magnetic boundary conditions.

For $\Lambda \gg 1$ the radiative zone
divides into two dynamically distinct regions,
separated by a thin internal boundary layer at $z=z_0$, say
(see figure \ref{E+Lambdavary}c).
The region above the boundary layer is close to Taylor--Proudman balance,
with all components of $\bu$ independent of $z$,
indicating that Lorentz forces are locally negligible
(see figure \ref{fig:uTwithsigma}a in \S\ref{sec:numexp_strat}
for plots of $\hux$).
In the region below the boundary layer, on the other hand,
the Lorentz force is dominant, and all components
of $\bu$ decay exponentially with depth.
From here on we refer to these two regions as being
``weakly magnetic'' and ``magnetically dominated'' respectively.
The boundary layer separating these two regions we call the
``magnetic transition layer''.
Since the flow within the magnetically dominated region decays
exponentially with depth, the solution is insensitive to
the choice of boundary conditions at $z=0$.

\begin{figure}[h]
\epsscale{1}
\plotone{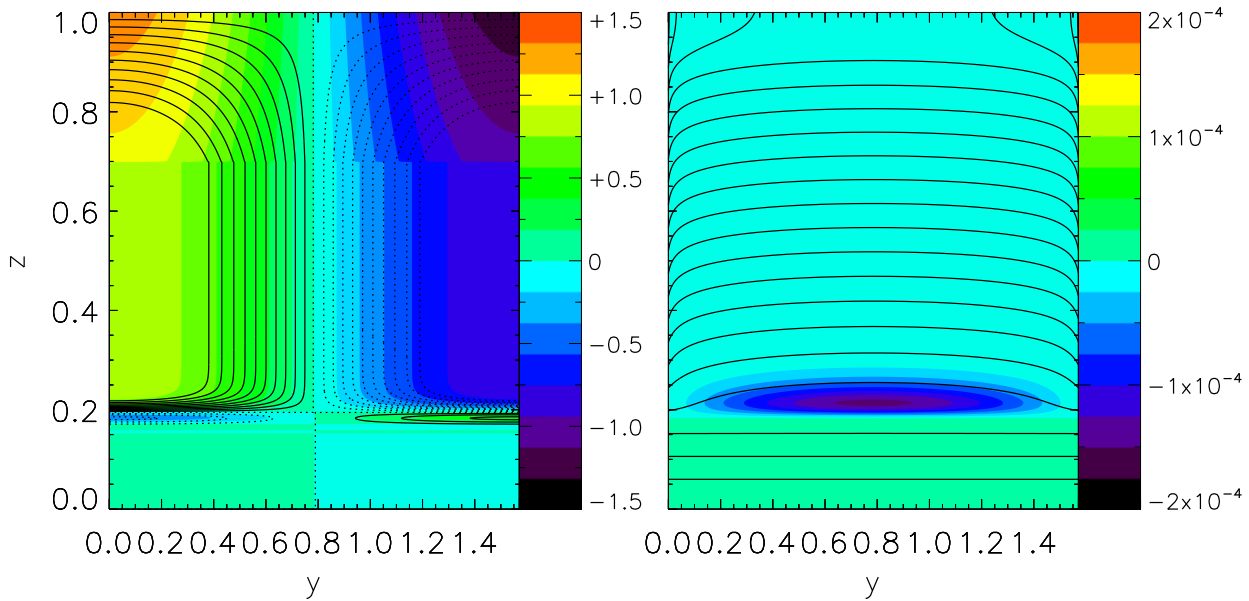}
\caption{\captionsize
  Meridional cross-section through the solution in figure
  \ref{E+Lambdavary} with $\Lambda = 10^{12}$ and
  $\Enu = 10^{-8}$. The left panel shows the azimuthal
  velocity in color, overlaid with selected meridional streamlines
  (solid lines for anticlockwise flow, dotted
  lines for clockwise flow).  The right panel shows
  lines of the poloidal magnetic field, including the background field $\bB_0$,
  assuming a magnetic Ekman number of $\Eeta = 10^{-4}$.
  Although field lines have been drawn at regular vertical intervals,
  the magnetic field strength decays exponentially with height.
  The color scale shows $\bx(x,y)/{\cal B}_0$.
  }
  \label{colorplots}
\end{figure}

The structure of the weakly magnetic and magnetically dominated
regions is illustrated further in figure \ref{colorplots},
which shows the meridional streamlines and magnetic field lines,
as well as contours of the azimuthal components of the velocity and magnetic
fields, in a vertical cross-section through a typical solution.
Within the convection zone, the forcing that drives differential rotation
also gyroscopically pumps a meridional flow.
Below the base of the convection zone the forcing is switched off, and
Taylor--Proudman balance is achieved. As a result,
the meridional streamlines and angular velocity contours become
aligned with the rotation axis, and a portion of the convection
zone's differential rotation extends into the radiative zone.
This differential rotation winds up the magnetic field lines, producing a
Lorentz torque whose amplitude increases exponentially with depth.
Within the magnetic transition layer (here around $z_0= 0.2$)
the Lorentz torque is strong enough to overcome Taylor--Proudman
balance, and gyroscopically pumps a meridional flow.
The magnetic field lines are dragged upward where the flow is
upwelling, $0 < y < \pi/2$, and pushed downward where the flow is
downwelling, $\pi/2 < y < \pi$ (not shown in the figure).
Below the transition layer,
meridional flows and differential rotation are both
strongly suppressed, and consequently perturbations
to the magnetic field lines are smaller.

\begin{figure}
\epsscale{0.5}
\plotone{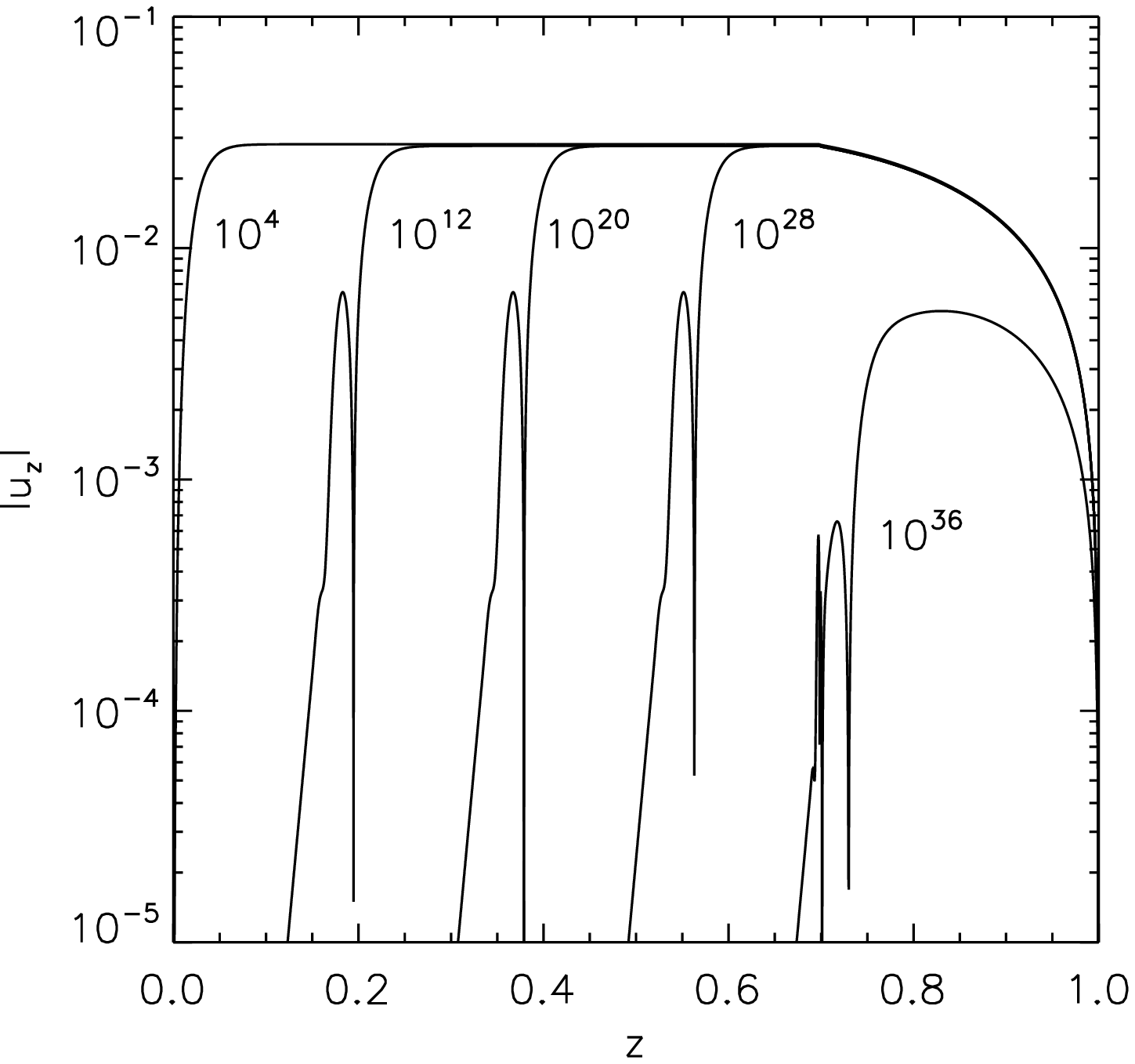}
  \caption{\captionsize
  Profiles of vertical velocity for a series of unstratified solutions
  with varying magnetic field strengths.  The Elsasser number $\Lambda$ was increased
  from $10^4$ to $10^{36}$ in multiplicative increments of $10^8$.
  Other parameters are as in figure~\ref{colorplots}.
  In each solution, the magnetic transition layer is located where the
  \emph{local} Elsasser number
  $\Lambda_{\rm loc} = \Lambda\ee^{-2z/\delta} \simeq 2/(k\delta)^2$
  (see \S\ref{transition-layer_unstrat}).
  }
  \label{Lambda_increase}
\end{figure}

Figure \ref{Lambda_increase} shows the result of increasing $\Lambda$ still further.
Over many orders of magnitude the only effect of increasing $\Lambda$ is to raise
the position of the magnetic transition layer, shrinking the
vertical extent of the
weakly magnetic region.
The dynamics within the convection zone remain unaffected until the point at which the
transition layer meets the base of the convection zone.

It is obviously of interest to identify the factors determining the location
of the magnetic transition layer.  We anticipate that
this transition
occurs
where the local Elsasser number $\Lambda_{\rm loc}$, defined in (\ref{eq:Lambdaeff}),
takes a critical value. We therefore expect the vertical position of the transition layer, $z_0$,
to follow a law of the form
\begin{equation}
  z_0 = \mbox{const.} + \tfrac{1}{2}\delta\ln{\Lambda}
  \label{eq:z0_guess}
\end{equation}
as $\Lambda$ is increased,
where the constant depends on the
aforementioned
critical value of $\Lambda_{\rm loc}$.
It is readily verified from figure \ref{Lambda_increase} that $z_0$ does indeed
follow such a law.
In the following sections we analyze the structure of the solutions in detail
in order to determine both the critical value of $\Lambda_{\rm loc}$
and the amplitude of the gyroscopically pumped meridional flow
below the convection zone.

\subsection{Analytical solution}
\label{analytics_unstrat}

In order to understand the results presented in \S\ref{sec:numexp_unstrat} more quantitatively, we now derive approximate analytical solutions of the governing equations.
In \S\ref{transition-layer_unstrat}
we derive a boundary-layer solution for the magnetic transition layer.
In \S\ref{sec:fullsol-unstrat}
we use that boundary-layer solution to construct a global analytical solution.

\subsubsection{Transition layer solution}
\label{transition-layer_unstrat}

We seek to describe the transition layer between the weakly magnetic
and magnetically dominated regions of the flow.  Since this transition occurs
below the base of the convection zone we may neglect the terms
in the momentum equation involving $\lambda$.
Based on the numerical solutions shown above,
we anticipate that the thickness of the transition layer
is of the same order as $\delta$, the scaleheight of the
background magnetic field.  Since
$\delta \ll 1/k$, we make the boundary-layer approximation
\begin{equation}
  \left(\frac{\dd^2}{\dd z^2} - k^2\right) \simeq \frac{\dd^2}{\dd z^2}\,.
  \label{eq:BLapprox}
\end{equation}
We also neglect viscosity, i.e., we set $\Enu = 0$.  It can be verified
\emph{a posteriori} that viscosity plays no
significant role in the dynamics of the transition layer
provided that $\delta$ is much larger than the Ekman
length, that is, provided that
\begin{equation}
  \delta \gg \Enu^{1/2}.
\end{equation}
This condition is easily satisfied in the solar interior, where $\Enu \simeq 10^{-15}$.  

With these approximations, the governing equations
(\ref{xmom})--(\ref{divb}) reduce to the following set:
\begin{eqnarray}
-2\huy &=& \ii k \Lambda \hbx\ee^{-z/\delta} \label{xmom2} \\
2\hux  &=& -\ii k\hat{p}-\frac{\Lambda}{\delta} \hbz\ee^{-z/\delta} \label{ymom2} \\
\frac{\dd\hat{p}}{\dd z} &=& \frac{\Lambda}{\delta}\hby\ee^{-z/\delta}
- \Lambda\frac{\dd\hby}{\dd z}\ee^{-z/\delta} \label{zmom2} \\
0 &=& \ii k\hux\ee^{-z/\delta} + \frac{1}{\delta}\frac{\dd\hbx}{\dd z} + \frac{\dd^2\hbx}{\dd z^2} \label{xind2} \\
0 &=& \ii k\huz\ee^{-z/\delta} + \frac{1}{\delta}\frac{\dd\hbz}{\dd z} + \frac{\dd^2\hbz}{\dd z^2} \label{intind2} \\
\ii k\huy +\frac{\dd\huz}{\dd z} &=& 0 \label{divu2} \\
\ii k\hby +\frac{\dd\hbz}{\dd z} &=& 0 \label{divb2}\,.
\end{eqnarray}
We show in appendix~\ref{sec:app-analytic} that
these can be combined into a single
equation for $\huz$:
\begin{equation}
  \left(\frac{\dd}{\dd z}-\frac{2}{\delta}\right)
    \left[
      \left({\rm e}^{z/\delta}\frac{\dd}{\dd z}\right)^4 + \tfrac{1}{4}k^4\Lambda^2
    \right]\huz  = 0\,.
  \label{eq:transition}
\end{equation}
From inspection of (\ref{eq:transition}) we deduce
that the vertical position of
the transition layer, $z_0$, is given by
\begin{align}
  ({\rm e}^{z_0/\delta}/\delta)^4 &= \tfrac{1}{4}k^4\Lambda^2 \nonumber \\
  \Rightarrow z_0 &= \delta \ln\left(k\delta/\sqrt{2}\right)
  + \tfrac{1}{2}\delta\ln{\Lambda}\,,
\label{eq:z0pred}
\end{align}
which is consistent with (\ref{eq:z0_guess}).
Furthermore, this expression shows that the transition layer is located where
the local  Elsasser number $\Lambda_{\rm loc}$, defined by (\ref{eq:Lambdaeff}),
takes the critical value $2/(k\delta)^2$.
This corresponds to a critical magnetic field strength
\begin{equation}
	B_{\rm crit} = \frac{\sqrt{8\pi\eta\rho_0\Omega_\odot}}{k\delta}
        \simeq 40 \left( \frac{\eta}{\eta_\odot} \right)^{1/2}
          \left(\frac{0.001}{\delta} \right) {\rm G} \,,
	\label{eq:Bcrit}
\end{equation}
where $\rho_0 \simeq 0.2$\,g\,cm$^{-3}$
and $\eta_\odot \simeq 400$\,cm$^2$s$^{-1}$. 

We seek the solution to (\ref{eq:transition})
that has the correct behavior in both the weakly magnetic region,
for $z - z_0 \gg \delta$, and the
magnetically dominated region, for $z - z_0 \ll -\delta$.
In particular, the solution must match onto the
differential rotation, $\hux  = u_{\rm t}$ say, in the
region immediately above the transition layer.  
Here, and subsequently, we use a subscript ``t'' to refer to the region
immediately above the transition layer,
which is also the bottom of the weakly magnetic region.
We show in appendix~\ref{sec:app-analytic} that the unique solution
with the required properties is
\begin{equation}
	\huz = \ii k\delta u_{\rm t}\left[\raisebox{0.5cm}{}
	  I_1(\zeta) - \tfrac{\pi}{2}\,{\rm Re}\left\{
	    \exp\left(-\tfrac{1+\ii}{\sqrt{2}}\zeta\right)
	  \right\}
  \right]\,, \label{eq:analytical-solution}
\end{equation}
where $\zeta = \exp((z_0-z)/\delta)$, and $I_1$ is the integral
\begin{equation}
	I_1(\zeta) = \int_0^\infty\!\ee^{-\zeta s}\frac{s\,\dd s}{1+s^4}\,.
\end{equation}
The weakly magnetic and magnetically
dominated regions correspond to $\zeta \ll 1$ and $\zeta \gg 1$ respectively.

\begin{figure}[h]
\epsscale{0.5}
\plotone{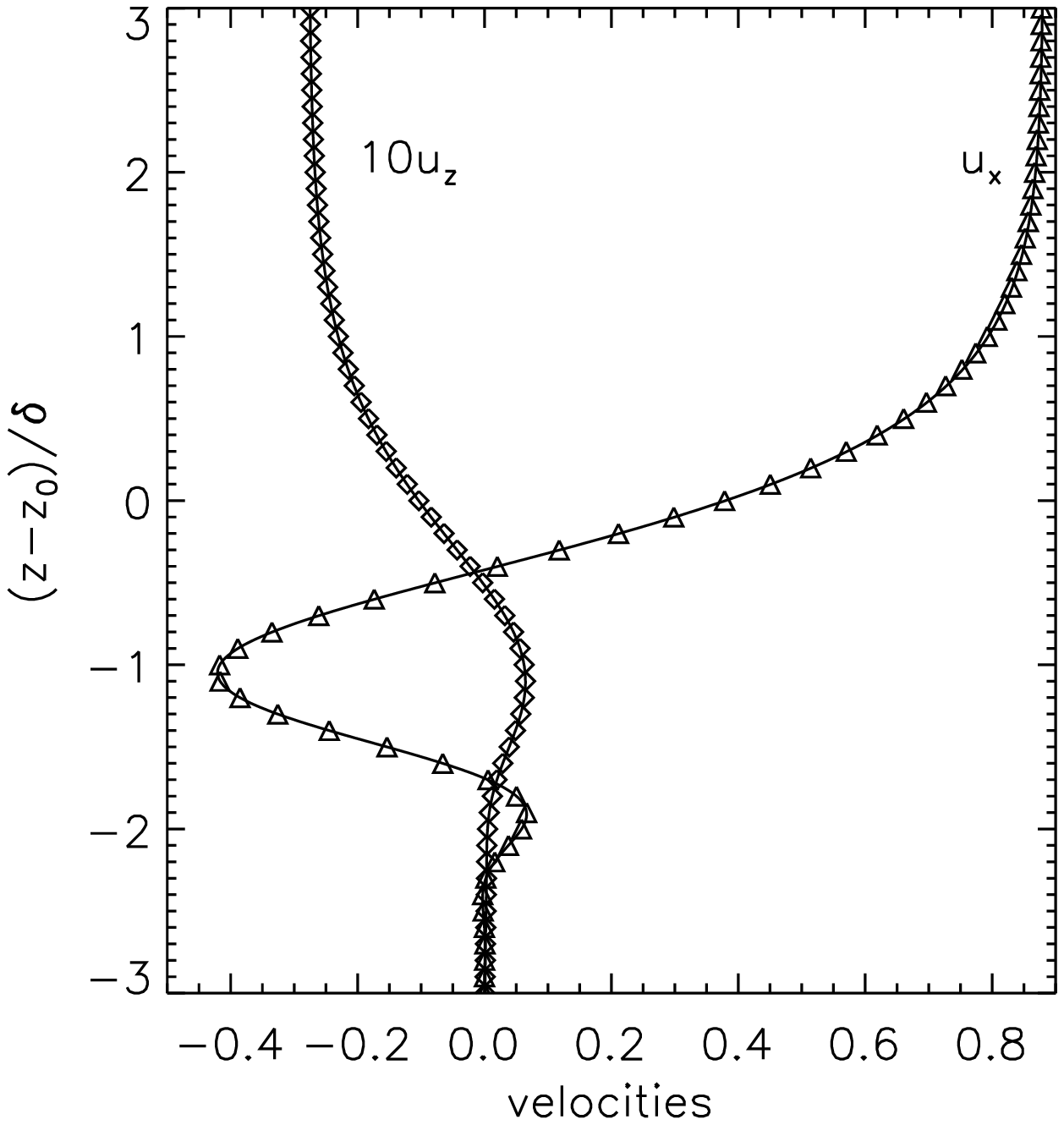}
\caption{\captionsize The vertical profiles of $\huz$ and $\hux$
  within the magnetic transition layer.  The solid lines correspond to the
  numerical solution shown in figure~\ref{colorplots};
  the symbols correspond to the analytical boundary-layer solution.
  The $\huz$ profile has been multiplied by 10 to make the two
  profiles visible on the same scale. The vertical axis represents
  the boundary-layer coordinate $(z-z_0)/\delta$, and so the convection zone lies
  far above the region plotted.
  }
  \label{fig:analytic}
\end{figure}

Figure~\ref{fig:analytic} shows the vertical profiles of $\huz$
and $\hux$, in the vicinity of the magnetic transition layer,
from the same numerical
solution shown in figure~\ref{colorplots}.
Also shown are their analytical counterparts,
(\ref{eq:analytical-solution}) and the corresponding analytical profile of $\hux$.

Within the magnetically dominated region,
where $\zeta \gg 1$,
the flow is exponentially weak.
From (\ref{eq:analytical-solution}) we find that
\begin{equation}
  \huz \sim \frac{\ii k\delta u_{\rm t}}{\zeta^{2}}
  = \ii k\delta u_{\rm t}\exp\left(\frac{2(z-z_0)}{\delta}\right).
  \label{eq:wmag}
\end{equation}
Within the weakly magnetic region,
where $\zeta \ll 1$,
$\huz $ tends to a
constant, $w_{\rm t}$ say, as expected 
from the Taylor--Proudman constraint.
Specifically,
\begin{equation}
  w_{\rm t} =
  \left.\huz\right|_{\zeta=0}
  = -\ii\tfrac{\pi}{4}k\delta\, u_{\rm t}.
  \label{eq:magnetic-pumping}
\end{equation}
The vertical mass flux within the weakly magnetic region is therefore
tied to the differential rotation via the magnetic transition layer.

\subsubsection{Global solution}
\label{sec:fullsol-unstrat}

We now use the relation (\ref{eq:magnetic-pumping})
derived from our transition-layer solution
to construct an approximate analytical solution for the
global flow.
In this way we derive an explicit relation between the
constant vertical velocity $w_{\rm t}$
within the weakly magnetic region
and the prescribed differential rotation $u_{\rm cz}(z)$
within the convection zone.
To construct the global solution,
we must also match the flow within the
weakly magnetic region, $z_0<z<h$, to the flow within the convection
zone, $h<z<1$. The matching procedure is similar to that employed by \citetalias{GaraudBodenheimer10},
and is described in appendix~\ref{sec:app-global}.
Here we present only the result for $w_{\rm t}$.

In the absence of stratification, $n_{\rm rz}=0$,
we find that $w_{\rm t}$ is given by
\begin{equation}
  w_{\rm t} \;\; = \;\;
  \dfrac{
    \raisebox{0.1cm}{$-\ii\bar{u}_{\rm cz}$}
  }{
    \dfrac{kd}{2\tau_c}\coth\left(\dfrac{1-h}{d}\right)
    + \dfrac{4}{\pi k\delta}
  }
\label{eq:w0estim}
\end{equation}
where $d$ is the ``convective'' lengthscale\footnote{The lengthscale $d$ in this paper is equivalent to the lengthscale $\delta_{\rm out}$ in \citetalias{GaraudBodenheimer10}.}
\begin{equation}
  d = \frac{\sqrt{4\tau_c^2 + 1}}{k}
  \label{eq:d}
\end{equation}
and where $\bar{u}_{\rm cz}$ is a weighted average
of the forcing in the convection zone,
\begin{equation}
  \bar{u}_{\rm cz}
  \;\;=\;\;
  \frac{\displaystyle\int_h^1\!\!\dd z\,u_{\rm cz}(z)\cosh\left(\frac{1-z}{d}\right)}
    {\displaystyle\int_h^1\!\!\dd z\,\cosh\left(\frac{1-z}{d}\right)}\;.
  \label{eq:average}
\end{equation}
Since $d > 1/k > (1-h)$, the weight distribution in (\ref{eq:average})
is roughly uniform, and so $\bar{u}_{\rm cz}$ is numerically
close to the vertical average of $u_{\rm cz}(z)$ over the convection zone.

\begin{figure}[h]
\epsscale{0.5}
\plotone{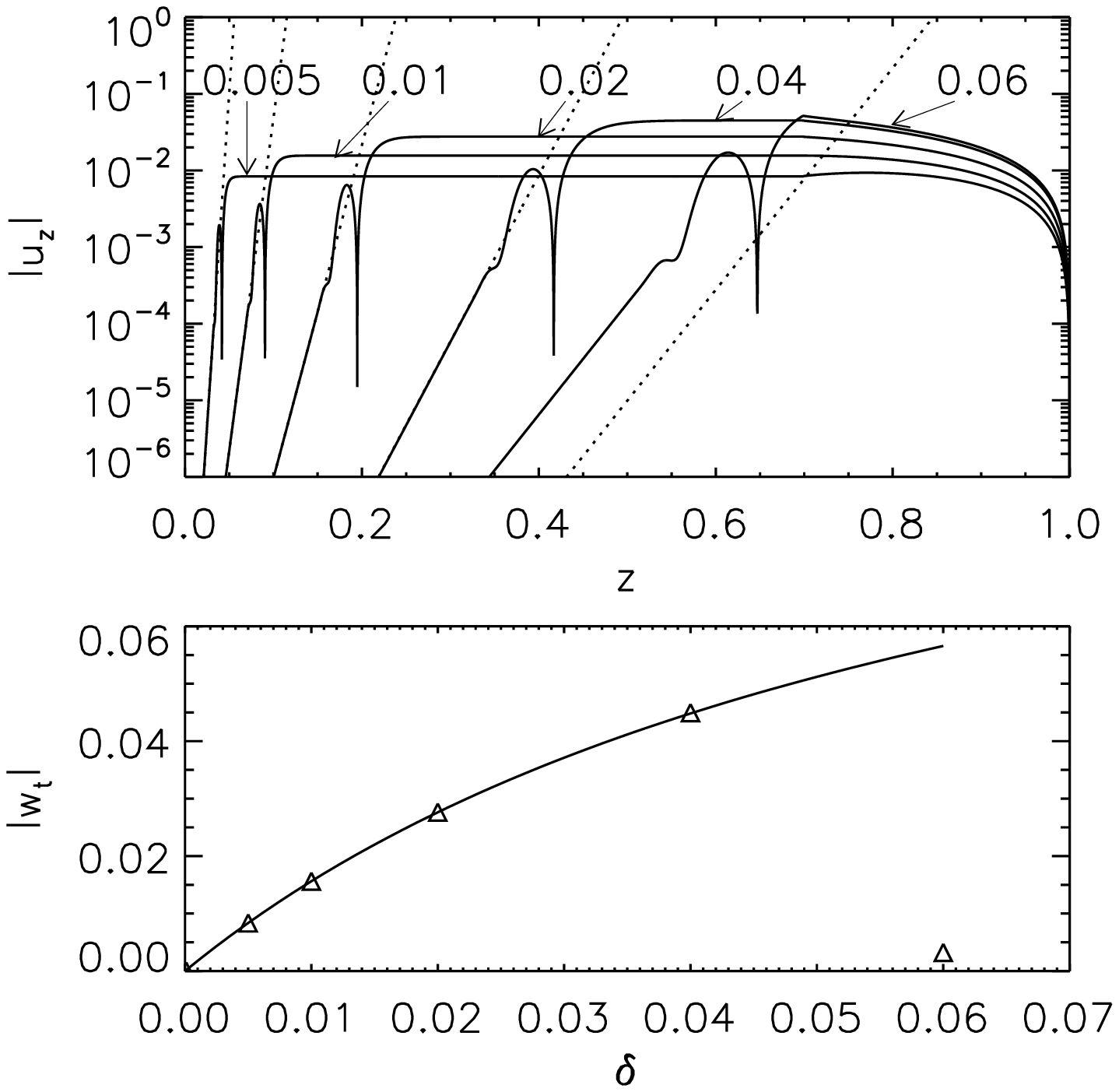}
  \caption{\captionsize
  Top panel: Vertical velocity profiles
  for various values of the magnetic scaleheight $\delta$.
  Other parameters are the same as in figure~\ref{colorplots}.
  The dotted lines show the asymptotic behavior (\ref{eq:wmag})
  of the boundary-layer solution (\ref{eq:analytical-solution})
  in the magnetically dominated region.
  Bottom panel: The vertical velocity $w_{\rm t}$ in the weakly
  magnetic region predicted by (\ref{eq:w0estim}).
  The symbols show the value of $\huz$ at $z=0.65$
  in each of the numerical solutions shown in
  the top panel.
  }
\label{fig:num-anal-unstrat}
\end{figure}

Figure~\ref{fig:num-anal-unstrat}
shows the vertical profiles of $\huz$ in a series of numerical solutions
with various magnetic scaleheights $\delta$.
The bottom panel in figure~\ref{fig:num-anal-unstrat}
compares the value of $\huz$ in the weakly magnetic region
to the value for $w_{\rm t}$ predicted by (\ref{eq:w0estim}).
The fit is excellent provided that
the magnetic transition layer is located sufficiently
far above the lower boundary, $z=0$, and sufficiently far
below the base of the convection zone, $z=h$.
Using (\ref{eq:z0pred}),
these two conditions can be expressed as
lower and upper bounds on $\Lambda$:
\begin{equation}
  \frac{2}{(k\delta)^2} \; \ll \; \Lambda \; \ll \;
  \frac{2{\rm e}^{2h/\delta}}{(k\delta)^2}.
\end{equation}

\subsection{Summary and discussion for the unstratified case}

We have found that a confined magnetic field is able to
halt the burrowing of
the convection zone's differential rotation
into the radiative interior, provided that the
magnetic field strength
exceeds a critical value $B_{\rm crit}$
given by (\ref{eq:Bcrit}).
Because the field strength
in our model
increases monotonically with depth,
the radiative zone
divides into an upper, weakly magnetic region,
wherein $B_0(z) \ll B_{\rm crit}$,
and a lower, magnetically dominated region, wherein $B_0(z) \gg B_{\rm crit}$.
In the weakly magnetic region
the flow is subject to the Taylor--Proudman constraint,
and the differential rotation $\ux$
is independent of $z$.
In the magnetically dominated region
the flow is subject to the Ferraro
constraint, which suppresses both differential rotation
and meridional flows. In our model
$B_0(z)$ increases exponentially with depth on a fixed lengthscale $\delta$,
and so the transition
between the weakly magnetic and magnetically dominated regions
occurs across a layer of thickness $\simeq\delta$,
within which $B_0 \simeq B_{\rm crit}$.

The transition layer
regulates the mass flux that downwells from the convection zone,
as expressed by equation (\ref{eq:magnetic-pumping}).
This is closely related to the mechanical constraint mentioned
in \S\ref{sec:intro}, and can be explained as follows.
The latitudinal differential rotation below the base of the convection zone extends downward until it meets the transition layer, where it winds
up the magnetic field lines, producing a Lorentz torque that overcomes
Taylor--Proudman balance and gyroscopically pumps a meridional flow.
In the steady state, the downwelling mass flux within the weakly
magnetic region must be exactly that demanded by the transition layer,
which is expressed by (\ref{eq:magnetic-pumping}).
An analogy can be made with the more familiar Ekman pumping that occurs in
Ekman layers.  Indeed, if the transition layer were replaced by an
artificial frictional, impenetrable and non-magnetic horizontal boundary,
then (\ref{eq:magnetic-pumping})
would be replaced
by the Ekman-pumping formula,
\begin{equation}
  w_{\rm t} = -\frac{1}{2}\ii k\dEk\, u_{\rm t}
\end{equation}
\citepalias{GaraudAA09},
where $\dEk = \Enu^{1/2}$ is the nondimensional Ekman-layer thickness.
In the solutions presented here we have $\delta \gg \dEk$, and so
the magnetic transition layer pumps much stronger meridional flows than
would
be pumped by
an artificial Ekman layer.

The solutions we have found bear many similarities to the tachocline
model originally proposed by \citet{GoughMcIntyre98}.
In particular, we have a differentially rotating
region below the base of the convection zone,
which we might call the tachocline,
and a thin magnetic boundary layer at the base of this region, which
\citeauthor{GoughMcIntyre98} called the ``tachopause''.  The entire region
below the tachopause is held in uniform rotation by the confined
magnetic field.
However, an obvious shortcoming of our unstratified solutions is that the
``tachocline'' has no vertical shear, quite unlike the solar
tachocline.  In the \citeauthor{GoughMcIntyre98} model,
the tachocline's vertical shear is explained by the presence of
stable stratification, via
the thermal-wind relation.
In the following sections we reintroduce stratification
into our model, in order to better approximate
conditions within the solar tachocline.

\section{The Stratified Case}
\label{sec:strat}

The importance of stratification in our model can be measured in terms
of the dimensionless parameter $n_{\rm rz}^2/\Ekappa$, which appears
on the left-hand side of (\ref{ener}).
In fact, it follows from equation (\ref{eq:thermal}) that
\begin{equation}
	\frac{n_{\rm rz}^2}{\Ekappa} = \Omega_\odot t_{\rm ES}\,,
\end{equation}
so $n_{\rm rz}^2/\Ekappa$ is precisely the dimensionless local
Eddington--Sweet timescale.
In what follows we also use $\sigma = n_{\rm rz}(\Enu/\Ekappa)^{1/2}$
as a measure of the stratification.  As mentioned in \S\ref{sec:intro},
this parameter is particularly relevant in cases where buoyancy and
viscosity both contribute to the balance of forces.
All the numerical solutions presented in this section have
$\Enu=10^{-8}$, and so the parameters
$n_{\rm rz}^2/\Ekappa$ and $\sigma$ contain equivalent information.

\subsection{Numerical exploration}
\label{sec:numexp_strat}

\begin{figure}[h!]
  \plotone{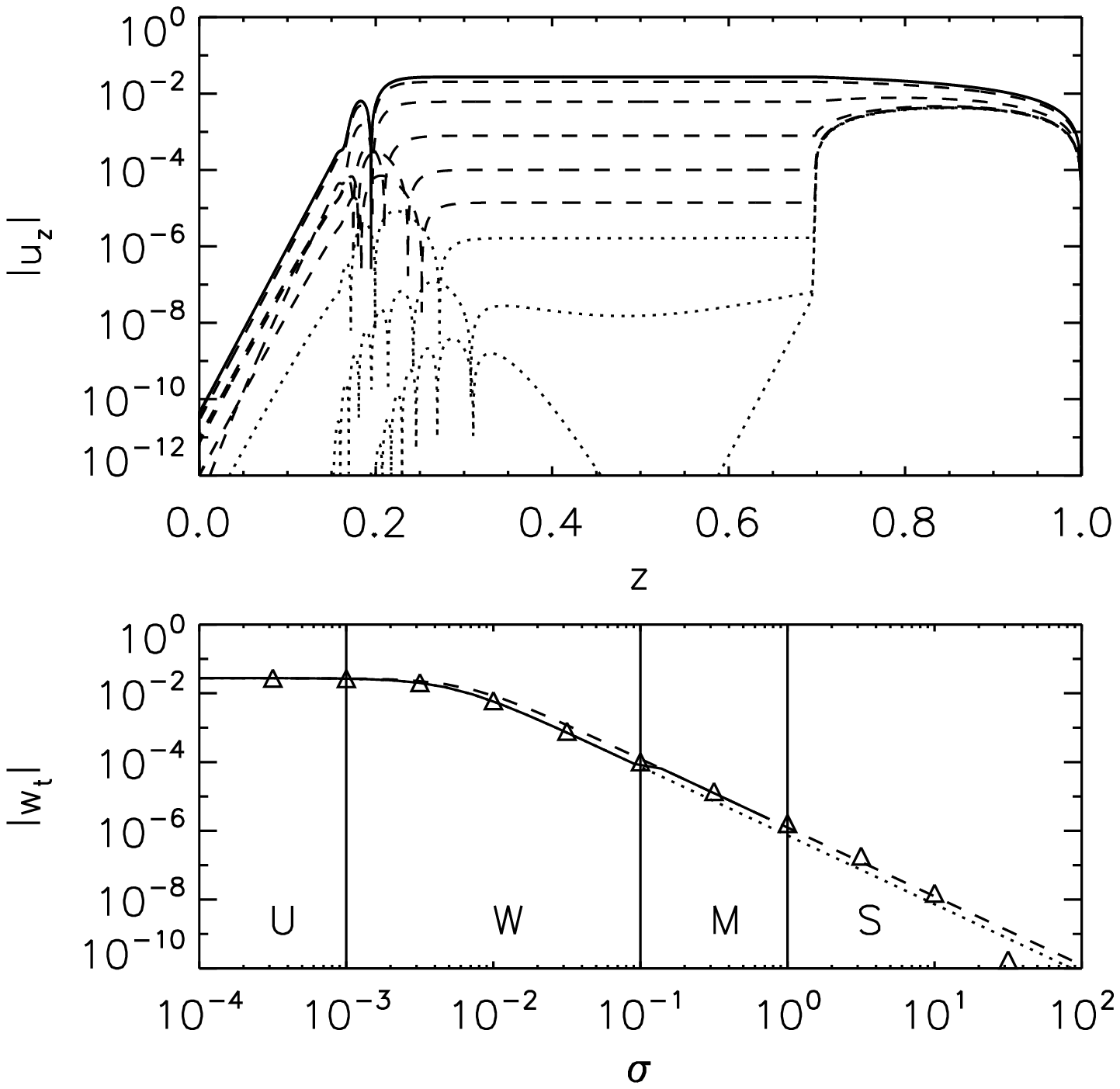}
  \caption{\captionsize
    Top panel: Profiles of $\huz(z)$ from a series of solutions
    with varying $\sigma$.
    All the solutions have $\fk = 10^6$ in (\ref{D}),
    and the other parameters are the same as in figure~\ref{colorplots}.
    We use solid lines for the unstratified solutions
    ($\sigma = 10^{-5},10^{-4},10^{-3}$),
    dashed lines for the weakly and moderately stratified solutions
    ($\sigma = 10^{-2.5},10^{-2},10^{-1.5},10^{-1},10^{-0.5}$),
    and dotted lines for the strongly stratified solutions
    ($\sigma = 10^{0},10^{1},10^{2}$).
    Bottom panel:
    The dotted and dashed lines show the vertical velocity $w_{\rm t}$
    in the tachocline
    predicted by (\ref{eq:w0estim2})
    for the weakly stratified and
    moderately stratified regimes respectively;
    a solid line is used to indicate the range of $\sigma$ values in which
    each is expected to apply.
    The symbols show the value of $\huz$ at $z=0.5$.
    The vertical lines indicate the boundaries between the unstratified,
    weakly stratified, moderately stratified and strongly stratified regimes.
    }
   \label{fig_3regimes}
\end{figure}

\begin{figure}[h]
\epsscale{1}
   \plottwo{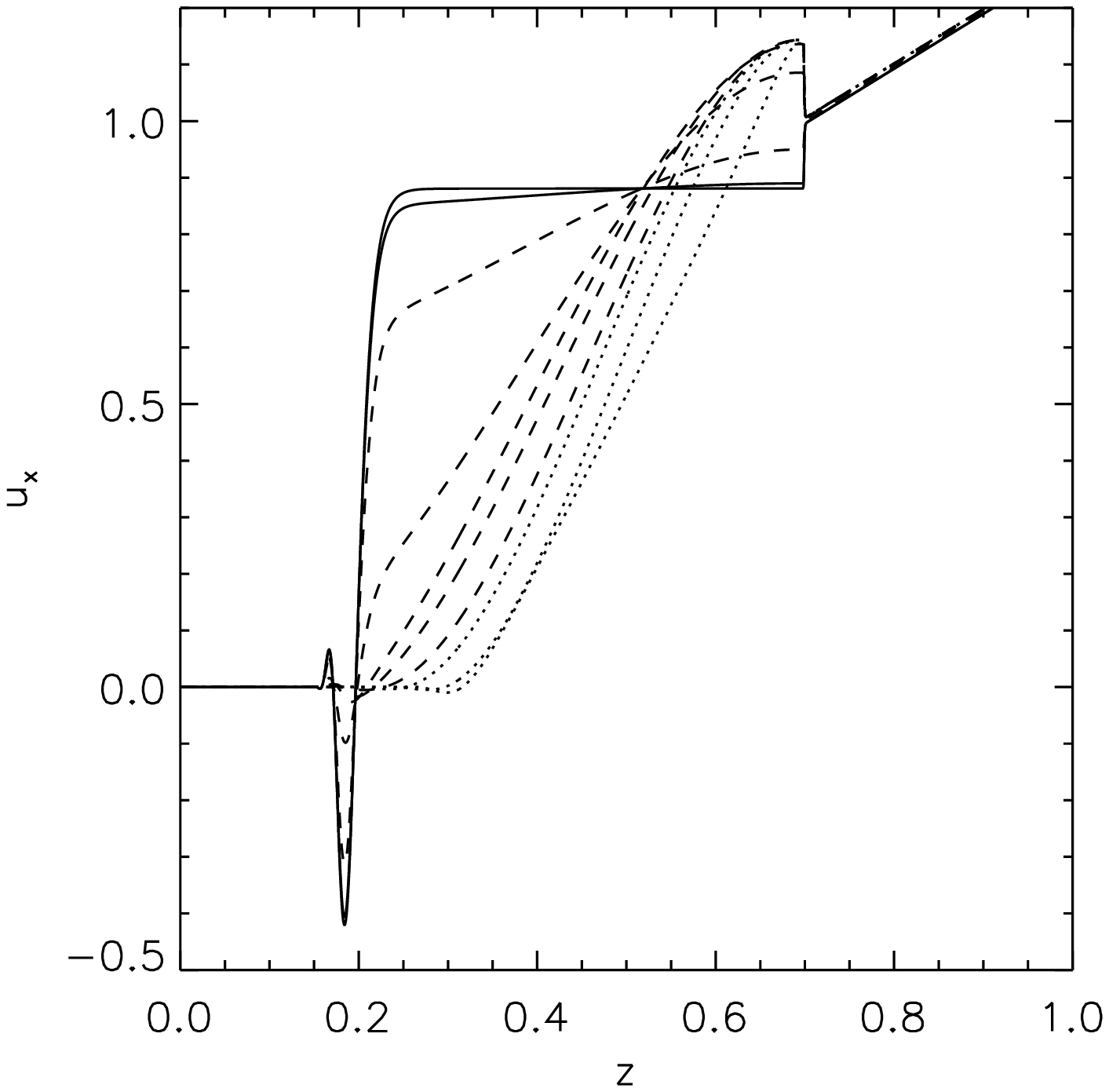}{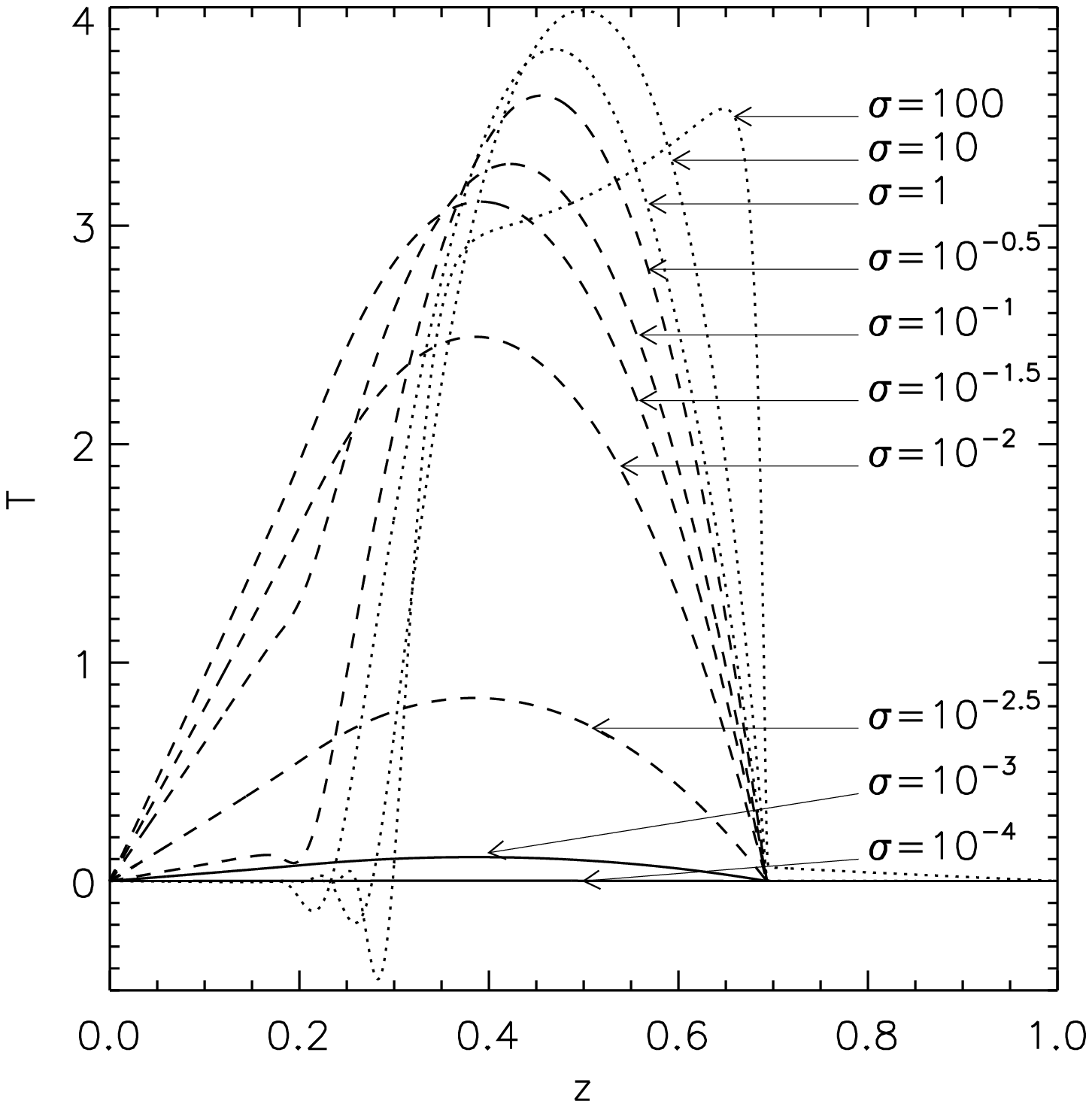}
   \caption{\captionsize
     Profiles of $\hux(z)$ and $\hat{T}(z)$ from the same solutions
     shown in figure~\ref{fig_3regimes}.
     The linestyles are the same as in figure~\ref{fig_3regimes}.
     Note that the vertical axes are linear.
   }
   \label{fig:uTwithsigma}
\end{figure}

We begin again by exploring the behavior of the solutions of (\ref{xmom})--(\ref{divb}) 
with varying governing parameters.
In figure \ref{fig_3regimes} we show
the vertical profiles of $\huz$ in
a series of numerical solutions
with increasing stratification, measured in terms of $\sigma$.
When $\sigma$ is sufficiently small,
$\lesssim 10^{-3}$,
the profiles are indistinguishable
from one another and from the unstratified solution, which has $\sigma=0$.
The same is true for the profiles of $\hux$ in these solutions,
which are shown in figure~\ref{fig:uTwithsigma}a.
In particular, the weakly magnetic region ($z_0 < z < h$)
remains in Taylor--Proudman balance.
We refer to all such cases as ``unstratified'', since any effects due to
stratification are negligible.

As $\sigma$ is increased beyond $10^{-3}$ we observe a reduction in
$\huz$
in the weakly magnetic
region
(see figure~\ref{fig_3regimes}).
At the same time, the jump in $\hux$ across the magnetic transition layer
drops almost to zero, and instead $\hux(z)$ rises smoothly and monotonically
between $z_0$ and $h$, indicating that thermal-wind balance has been established
within the weakly magnetic region.
For brevity, and because of the obvious similarity with the model of
\citet{GoughMcIntyre98}, from here on we refer to the weakly magnetic region and
magnetic transition layer in such cases as the ``tachocline'' and
``tachopause'' respectively.

We find that cases with significant stratification
can be further categorized as being either ``weakly stratified'', ``moderately stratified'',
or ``strongly stratified''.
These three regimes are now described in turn.

\subsubsection{The strongly stratified regime}

In common with \citet{GaraudBrummell08} and
\citetalias{GaraudAA09}, we find that the strength of the meridional flow
$\huz$
decays exponentially with depth below the convection zone
on the vertical scale $1/(k\sigma)$.
If $\sigma$ is sufficiently large, $\gtrsim 1$, then
this vertical scale is smaller than the tachocline
thickness, and so meridional flows gyroscopically pumped
at the base of the convection
zone are halted by a combination of buoyancy and viscous forces
before reaching the tachopause.
Following the terminology introduced by \citetalias{GaraudAA09}
we refer to such cases as ``strongly stratified''.
Even with strong stratification,
the differential rotation $\hux$ driven in the convection zone
still spreads into the radiative interior (figure~\ref{fig:uTwithsigma}a),
but by viscous diffusion rather than by the burrowing of meridional circulations.
The differential rotation
winds up the magnetic field lines
much as in the unstratified cases described in \S\ref{sec:unstrat},
producing a Lorentz torque within the tachopause that gyroscopically pumps
a meridional flow.
Part of
that meridional flow is launched upward into the tachocline,
where its strength decays with height on the lengthscale $1/(k\sigma)$.
The strongly stratified solutions therefore have
meridional circulations within two distinct horizontal layers,
at the top and bottom of the tachocline.

\subsubsection{The weakly stratified and moderately stratified regimes}
\label{sec:weak_and_mod}

Figure~\ref{fig:uTwithsigma}b shows vertical profiles of $\hat{T}$
from the same series of solutions shown in figure~\ref{fig_3regimes}.
In all cases with $\sigma \lesssim 10^{-1}$ we find that
there is little variation in the temperature field $\hat{T}$
or its vertical gradient $\partial\hat{T}/\partial z$
across the tachopause.
However, in cases with $\sigma \gtrsim 10^{-1}$
we observe a significant change in the temperature gradient
at $z=z_0$,
indicating that the flow within the tachopause contributes
to the global-scale
thermal equilibrium.
We refer to cases with $10^{-3} \lesssim \sigma \lesssim 10^{-1}$
as ``weakly stratified'', and cases with $10^{-1} \lesssim \sigma \lesssim 1$
as ``moderately stratified''.

\begin{figure}[h!]
\epsscale{0.7}
  \plotone{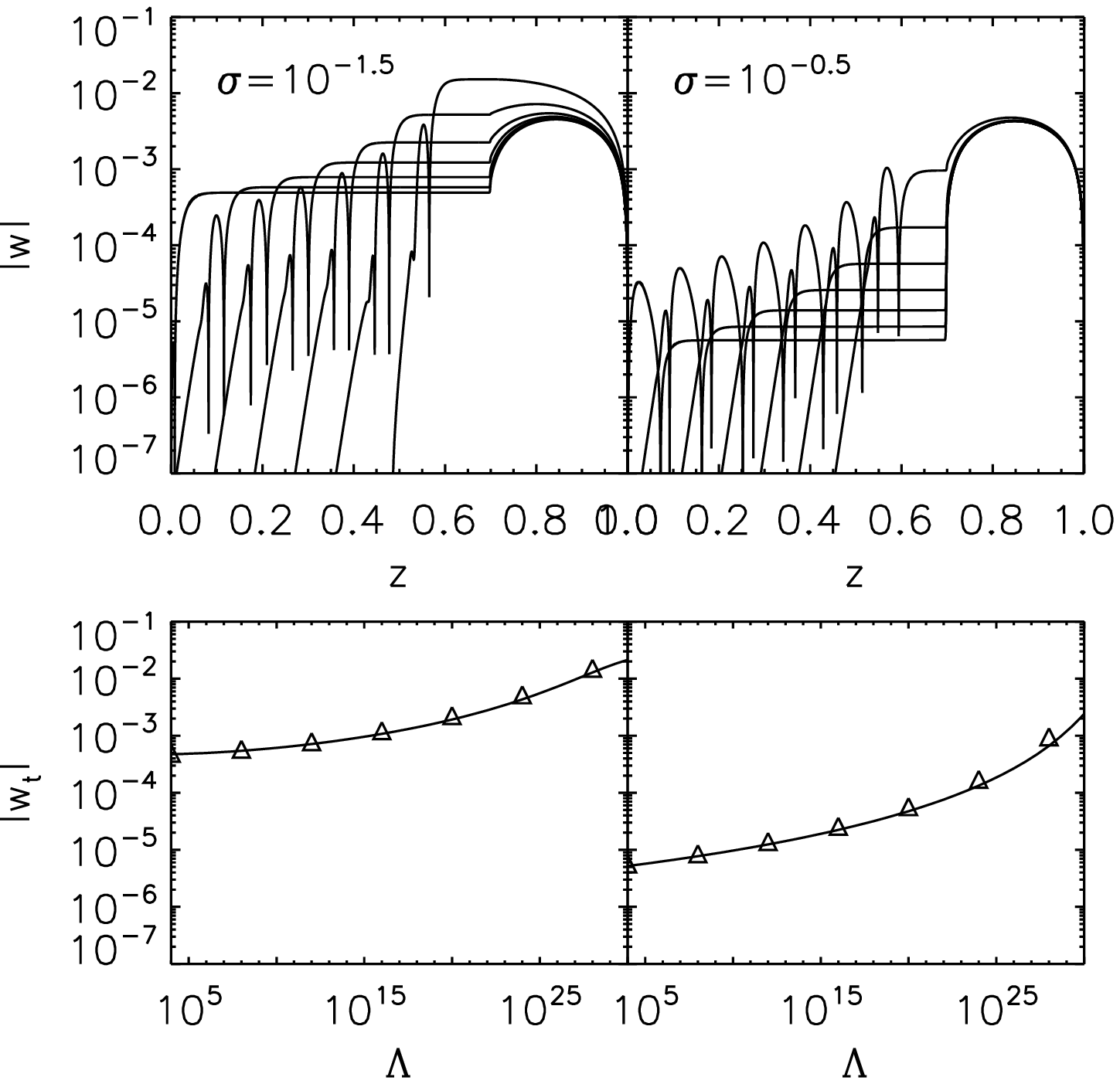}
  \caption{\captionsize
    Top panel: Profiles of $\huz(z)$ from a series of
    simulations with $\Lambda$ increasing from $\Lambda=10^{4}$ to
    $10^{28}$ in multiplicative increments of $10^{4}$,
    for $\sigma=10^{-1.5}$ (left column) and $\sigma=10^{-0.5}$ (right column).
    Other parameters are as in figure~\ref{fig_3regimes}.
    Bottom panel: The vertical velocity $w_{\rm t}$ in the tachocline
    predicted by (\ref{eq:w0estim2}) in the weakly stratified regime (left column)
    and moderately stratified regime (right column).
    The symbols show the value of $\huz$ at $z=0.65$, just below
    the base of the convection zone, for each of the profiles in
    the top panel.
  }
   \label{fig:varyb}
\end{figure}

Figure \ref{fig:varyb} illustrates how the mass flux in the tachocline
varies with the strength of the interior magnetic field
in the weakly and moderately stratified regimes.
As in the unstratified case, we find that the position of the
magnetic transition layer moves upwards as $\Lambda$ increases, in
a manner which is more-or-less independent of $\sigma$.
However, in contrast with the unstratified case,
the strength of the meridional flow within the tachocline
varies with $\Lambda$
(cf.~figure \ref{Lambda_increase}).
For increasing values of $\Lambda$, the meridional flow becomes
stronger as the tachocline becomes thinner.
This result has implications for the solar tachocline, as will be
discussed in \S\ref{sec:global-strat}.

\begin{figure}[h]
\epsscale{0.7}
  \plotone{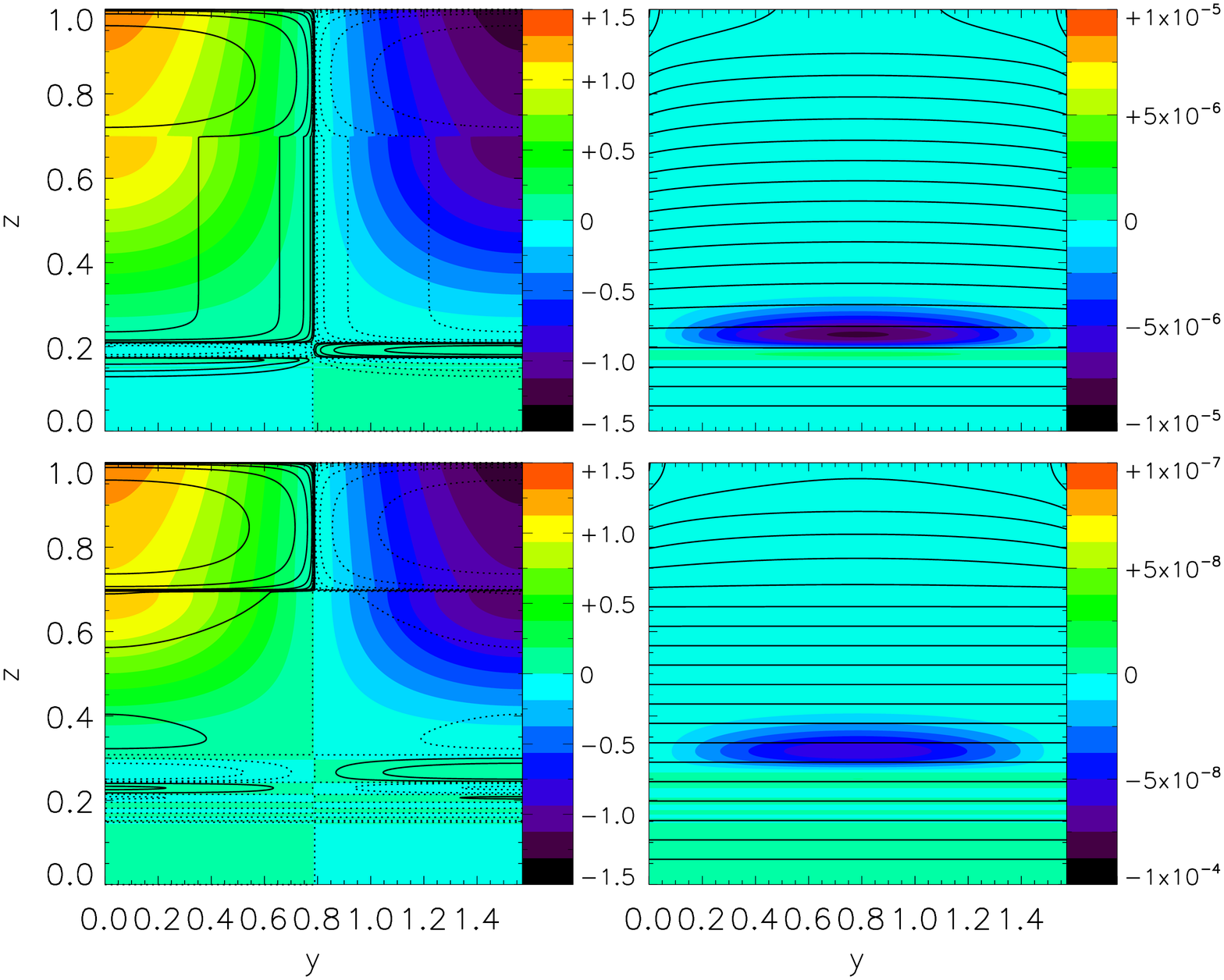}
  \caption{\captionsize
    Meridional cross-sections through the solutions in figure
    \ref{fig_3regimes} with $\sigma = 10^{-1.5}$ and
    $\sigma = 10^{2}$, corresponding to the weakly and strongly
    stratified regimes respectively. The left column shows the azimuthal
    velocity and meridional streamlines.  The right column shows
    the azimuthal magnetic field and poloidal field lines.
    Different streamlines have been plotted in the two cases
    because in the strongly stratified regime the meridional flow
    within the tachocline
    is weaker by many orders of magnitude.
  }
  \label{fig:colorplots2}
\end{figure}

Figure~\ref{fig:colorplots2} shows meridional cross-sections
through two solutions in the weakly and strongly stratified regimes.
As mentioned in \S\ref{sec:intro}, the solar tachocline
has $\sigma \simeq 0.2$--$0.4$ \citepalias{GaraudAA09},
and therefore is not strongly
stratified in the sense used here.
We anticipate that either the weakly stratified or the moderately stratified
regime offers the best approximation to the dynamics of the tachocline.
In the following sections we analyze
these two regimes in detail in order to
quantify more precisely the parameter ranges over which they apply.

\subsection{Analytical solution}
\label{analytics_strat}

We proceed, as in \S\ref{analytics_unstrat}, by deriving an analytical boundary-layer
solution for the tachopause, which can then be used
to construct a global analytical solution.
In order to simplify the analysis
we will neglect the viscous terms in the equations, and so our
analytical solutions cannot be applied in the strongly stratified regime.

\subsubsection{The tachopause}
\label{transition-layer_strat}

In the unstratified, weakly stratified and moderately stratified
solutions presented in figure~\ref{fig:uTwithsigma}b,
the temperature field $\hat{T}(z)$ is
roughly constant in
the region $z-z_0 = O(\delta)$.
(Recall that the moderately stratified cases are characterized
by a jump in $\dd\hat{T}/\dd z$, rather than $\hat{T}$.)
We can therefore analyze the structure of the tachopause
in all three regimes simultaneously
by approximating
$\hat{T}$ as a constant, $T_{\rm t}$.
As in \S\ref{transition-layer_unstrat} we neglect viscosity
and make the boundary-layer approximation (\ref{eq:BLapprox}).
We then recover equations
(\ref{xmom2})--(\ref{divb2}) exactly,
except that (\ref{zmom2}) is replaced by
\begin{equation}
	\frac{\dd\hat{p}}{\dd z} - T_{\rm t} \:=\:
	\frac{\Lambda}{\delta}\hby\ee^{-z/\delta}
	- \Lambda\frac{\dd\hby}{\dd z}\ee^{-z/\delta}\,.
	\label{zmom3}
\end{equation}
Within the tachocline,
where $z-z_0 \gg \delta$, all the terms involving
$\Lambda$ are negligible, and so we have
\begin{eqnarray}
  \huy &\simeq& 0 \\
  2\hux &\simeq& -\ii k\hat{p} \\
  \frac{\dd\hat{p}}{\dd z} - T_{\rm t} &\simeq& 0 \\
  \ii k\huy + \frac{\dd\huz}{\dd z} &\simeq& 0\,.
\end{eqnarray}
We deduce that
\begin{equation}
  \left.
  \begin{array}{rcl}
    \hux &\sim& u_{\rm t} - \tfrac{1}{2}\ii kT_{\rm t}(z-z_0) \\
    \huy &\to& 0 \\
    \huz &\to& w_{\rm t}
  \end{array}
  \right\} \;\;\; \mbox{as} \;\;\; \frac{(z-z_0)}{\delta} \to +\infty\,,
  \label{eq:match}
\end{equation}
where the constants $u_{\rm t}$
and $w_{\rm t}$ represent the values of $\hux$ and $\huz$
immediately above the tachopause, or equivalently,
at the bottom of the tachocline.

\begin{figure}[h]
\epsscale{0.5}
\plotone{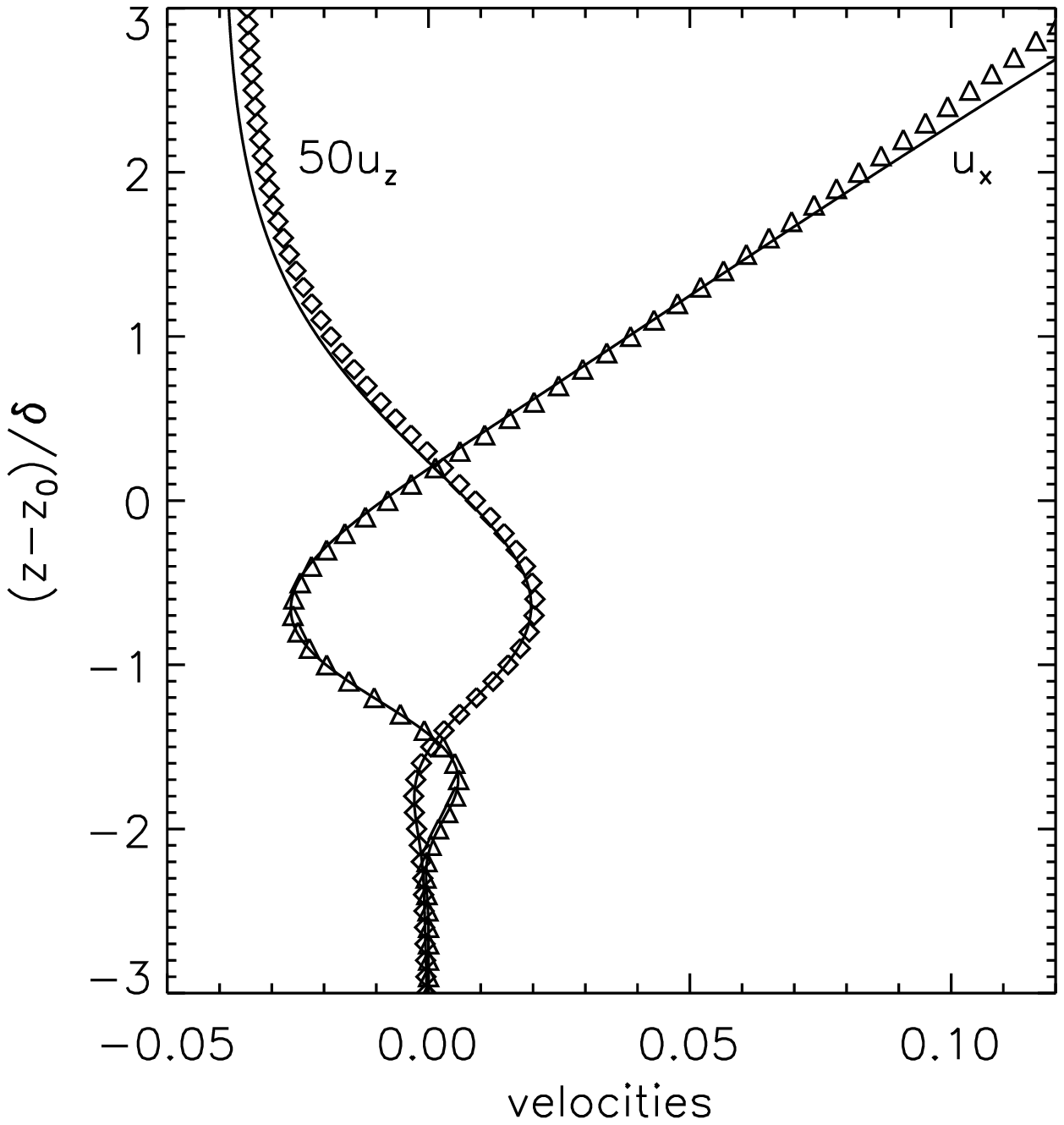}
\caption{\captionsize The vertical profiles of $\huz$ and $\hux$
  within the tachopause.  The solid lines correspond to the
  numerical solution shown in the top panel of figure~\ref{fig:colorplots2};
  the symbols correspond to the analytical boundary-layer solution.
  The $\huz$ profile has been multiplied by 50 to make the two
  profiles visible on the same scale.
  }
  \label{fig:analytic2}
\end{figure}

As in \S\ref{transition-layer_unstrat} we can
combine all of our boundary-layer equations into
a single equation for $\huz$:
\begin{equation}
  \left(\frac{\dd}{\dd z}-\frac{2}{\delta}\right)
    \left[
      \left(\ee^{z/\delta}\frac{\dd}{\dd z}\right)^4 + \tfrac{1}{4}k^4\Lambda^2
    \right]\huz = \frac{k^4\Lambda}{4\delta}T_{\rm t}\,\ee^{2z/\delta}\,.
  \label{eq:transition2}
\end{equation}
In appendix~\ref{sec:app-analytic} we show that the unique solution of
(\ref{eq:transition2}) that satisfies (\ref{eq:match}) is
\begin{align}
	\huz &= \ii k\delta u_{\rm t}\left[\raisebox{0.5cm}{}
	  I_1(\zeta) - \tfrac{\pi}{2}\,{\rm Re}\left\{
	    \exp\left(-\tfrac{1+\ii}{\sqrt{2}}\zeta\right)
	  \right\}
  \right] \nonumber \\
  &\hspace{1cm}+
  \tfrac{1}{2}k^2\delta^2 T_{\rm t}\left[\raisebox{0.5cm}{}
	  I_2(\zeta) - \tfrac{\pi}{2}\,{\rm Re}\left\{
	    (\gamma-\ii\tfrac{\pi}{4})\exp\left(-\tfrac{1+\ii}{\sqrt{2}}\zeta\right)
	  \right\}
  \right]
  \label{eq:analytical-solution2}
\end{align}
where $\gamma$ is the Euler--Mascheroni constant, $\gamma = 0.577...$,
$\zeta = \exp((z_0-z)/\delta)$, as in \S\ref{transition-layer_unstrat}, and
\begin{align}
  I_1(\zeta) &= \int_0^\infty\!\frac{s\,\dd s}{\ee^{\zeta s}}\,\frac{1}{1+s^4}\,, \\
  I_2(\zeta) &= \int_0^\infty\!\frac{s\,\dd s}{\ee^{\zeta s}}\,\frac{\gamma+\ln s}{1+s^4}\,.
\end{align}
Figure~\ref{fig:analytic2} shows the vertical profiles of $\huz$ and $\hux$,
and their analytical counterparts,
from the same numerical
solution shown in the top panel of figure~\ref{fig:colorplots2}.

From (\ref{eq:analytical-solution2})
we deduce a relation between the
constants $T_{\rm t}$, $u_{\rm t}$, and $w_{\rm t}$,
\begin{equation}
  w_{\rm t} = \left.\huz\right|_{\zeta=0}
  = -\frac{\ii\pi}{4}k\delta u_{\rm t}
    - \frac{\gamma\pi}{8}k^2\delta^2 T_{\rm t}\,.
  \label{eq:magnetic-pumping2}
\end{equation}
It can be verified that (\ref{eq:magnetic-pumping2}) holds, to reasonable
accuracy, in each of the solutions presented in \S\ref{sec:numexp_strat}.
Furthermore, in each case the term in (\ref{eq:magnetic-pumping2})
involving $T_{\rm t}$ is found to be negligible, and so (\ref{eq:magnetic-pumping2})
reduces to the relation (\ref{eq:magnetic-pumping}) derived for the
unstratified transition layer.
The vertical mass flux within the tachocline is therefore tied to the
differential rotation much as in the unstratified case described
in \S\ref{sec:unstrat}.

We can also use (\ref{eq:analytical-solution2}) to quantify the
role of the tachopause in the global-scale heat flow.
Within the tachopause the thermal energy equation (\ref{ener}) becomes
\begin{align}
	\frac{n_{\rm rz}^2}{\Ekappa}\huz = \frac{\dd^2\hat{T}}{\dd z^2}
\end{align}
after making the boundary-layer approximation (\ref{eq:BLapprox}).
Therefore the change in the vertical temperature gradient
across the tachopause is
\begin{align}
	\left[\frac{\dd\hat{T}}{\dd z}\right]_{z=z_0-\delta}^{z=z_0+\delta}
	\;\; &= \;\;
	\frac{n_{\rm rz}^2}{\Ekappa}\int_{z_0-\delta}^{z_0+\delta}\huz\,\dd z
	\nonumber \\
	\;\; &\sim \;\; \frac{n_{\rm rz}^2}{\Ekappa}w_{\rm t}\delta\,.
	\label{eq:heating}
\end{align}
The transition
between the weakly stratified and moderately stratified regimes
occurs when (\ref{eq:heating}) is of comparable magnitude to
the temperature gradient within the tachocline
(see figure~\ref{fig:uTwithsigma}b).

\subsubsection{Global solution}
\label{sec:global-strat}

We can use the relations (\ref{eq:magnetic-pumping2}) and (\ref{eq:heating})
to construct an approximate analytical solution for the global flow.
The details of the solution procedure are set out in
appendix~\ref{sec:app-global}.
We find that the vertical velocity $w_{\rm t}$ in the tachocline is
\begin{equation}
  w_{\rm t} \;\; = \;\; \dfrac{\raisebox{0.2cm}{$-\ii\bar{u}_{\rm cz}$}}
    {\dfrac{kd}{2\tau_c}\coth\left(\dfrac{1-h}{d}\right) + \dfrac{4}{\pi k\delta}
      + \dfrac{n_{\rm rz}^2}{2k^2\Ekappa}[G_1 + G_2 - G_3]}\,,
  \label{eq:w0estim2}
\end{equation}
which reduces to equation (\ref{eq:w0estim}) in the absence of
stratification, $n_{\rm rz} = 0$.
The dimensionless ``geometrical'' factors $G_1$, $G_2$ and $G_3$
are given by (\ref{eq:G1-weak})--(\ref{eq:G3-weak}) in the weakly stratified regime
and by (\ref{eq:G1-mod})--(\ref{eq:G3-mod}) in the moderately stratified regime.
We refer to these factors as geometrical
because of their dependence on the tachocline thickness, $D=h-z_0$.
However, they also depend on $\fk$, the thermal diffusivity enhancement factor
defined in (\ref{D}).
Moreover, $G_2$ depends on the lengthscale $d$ defined by (\ref{eq:d}),
and hence on the forcing timescale $\tau_c$.

We can use (\ref{eq:w0estim2}) to quantify
the boundary between the unstratified and weakly stratified regimes
identified in \S\ref{sec:numexp_strat}.
This boundary is
located where the term in the denominator of (\ref{eq:w0estim2})
involving $n_{\rm rz}$ becomes as large as the term involving $\delta$.
In general the geometrical factors are of order unity or smaller,
and so the unstratified regime corresponds to
\begin{align}
	\frac{n_{\rm rz}^2}{\Ekappa} \lesssim k/\delta\,.
	\label{eq:unstrat}
\end{align}
For the solutions in figures \ref{fig_3regimes} and \ref{fig:uTwithsigma},
which have $\Enu=10^{-8}$, $k=2$, and $\delta=0.02$, this condition
is equivalent to $\sigma \lesssim 10^{-3}$, in good agreement with the
numerical results.

The boundary between the weakly and
moderately stratified regimes occurs where the change in the temperature
gradient across the tachopause, given by (\ref{eq:heating}),
becomes comparable to the temperature gradient within the tachocline.
As described in appendix~\ref{sec:app-global}, this condition can be
expressed approximately as
\begin{align}
	\frac{n_{\rm rz}^2}{\Ekappa} \simeq 1/(k\delta^3)\,.
	\label{eq:weak_mod}
\end{align}
For the solutions in figures \ref{fig_3regimes} and \ref{fig:uTwithsigma}
this condition is equivalent to $\sigma \simeq 0.025$, which is
consistent with
(although somewhat smaller than)
the value $\sigma \simeq 0.1$ found in the numerical solutions.

The bottom panels of figures~\ref{fig_3regimes} and \ref{fig:varyb}
verify that the analytical prediction (\ref{eq:w0estim2})
for the vertical flow velocity in the tachocline
agrees with the numerical results described earlier.
Since viscosity was neglected in the derivation of (\ref{eq:w0estim2})
this good agreement demonstrates that the weakly stratified and
moderately stratified regimes are both inviscid, that is,
viscous forces do not play a significant role in the dynamics.
In particular, the solutions do not have an Ekman layer at the
radiative--convective interface, $z=h$.
We believe that the existence of such an Ekman layer in
some
previous studies
\citep[e.g.][]{GilmanMiesch04,Rudiger-etal05,GaraudBrummell08}
arises from the treatment of the interface in those studies.
In all the results presented
here, the change in buoyancy frequency $n$ at the interface is smoothed
over a lengthscale $\Delta$, representing the
depth of convective overshoot,
that greatly exceeds the Ekman length $\dEk = \Enu^{1/2}$.
In the other studies just mentioned,
the interface was modelled either as an upper boundary with
a fixed differential rotation,
or by a change in $n$ over a lengthscale $\ll \dEk$.
Although the precise thickness of the Sun's
overshoot layer is not known, it is surely thicker than the Ekman length
$(\nu/\Omega_\odot)^{1/2} \simeq 30$\,m, and so we argue that there is
no Ekman layer in the solar tachocline.

\subsection{Physical interpretation and discussion of the stratified results}
\label{sec:interpretation}
It is useful to compare the expression for $w_{\rm t}$
given by (\ref{eq:w0estim2})
to the corresponding expression derived by \citetalias{GaraudBodenheimer10}
for meridional circulations pumped
between the outer and inner convective zones
of young lithium-dip stars (see their equation (23)).
As in their model, the magnitude of the
circulation induced by the convective stresses is
proportional to $\bar u_{\rm cz}$, a weighted average of the
prescribed differential rotation in the convection zone.  
It is then moderated by whichever term in the denominator of
(\ref{eq:w0estim2}) is largest.
The first term is always of order unity or larger, and so
$w_{\rm t}$ can never exceed $\bar u_{\rm cz}$ in magnitude.
The second term, involving $\delta$, limits $w_{\rm t}$
to a flow rate which can
be accommodated by the magnetic transition layer, as discussed in
\S\ref{sec:unstrat}; this is an example of
the ``mechanical'' constraint mentioned in \S\ref{sec:intro}.
The third term, involving $n_{\rm rz}$, represents the thermal constraint
described in \S\ref{sec:intro}.
This constraint limits $w_{\rm t}$ to a
flow rate for which the temperature perturbations created by
the advection of the background stratification can be balanced by
thermal diffusion, thereby maintaining thermal equilibrium.

The expression for $w_{\rm t}$ given by (\ref{eq:w0estim2})
can be simplified considerably
in the parameter
regime appropriate to the solar interior.
In the solar tachocline we have $n_{\rm rz}^2/\Ekappa \simeq 8\times10^{13}$,
and so the tachocline is not in the unstratified regime according to the
criterion (\ref{eq:unstrat}) unless the tachopause is unrealistically thin,
$\delta \lesssim 10^{-13}$.
Therefore the denominator of (\ref{eq:w0estim2})
is dominated by the thermal term.
The weakly stratified and moderately stratified regimes,
according to (\ref{eq:weak_mod}),
correspond to $\delta \lesssim 10^{-5}$
and $\delta \gtrsim 10^{-5}$ respectively.
The solar tachocline could plausibly be in either of these regimes.
Fortunately, it can be shown that $G_1$, $G_2$ and $G_3$
follow similar scaling laws in
either case, differing only by factors of order unity.
We can therefore consider both regimes simultaneously.

We expect the tachopause to
be much thinner than the tachocline, $\delta \ll D$,
which is itself much thinner than the solar radius, $D \ll 1$.
Under these conditions
it can be shown that
\begin{flalign}
	&& G_1/G_2 &\sim \fk(kD)^2& \label{eq:G1_G2} \\
	&\mbox{and}& G_1/G_3 &\sim D/\delta \gg 1&
\end{flalign}
in both the weakly and moderately stratified regimes.
The right-hand side of (\ref{eq:G1_G2}) corresponds to
the ratio of the thermal adjustment timescales
in the tachocline and convection zone respectively.
In the solar tachocline, thermal adjustment
by radiative diffusion has a timescale of
about $10^4$ years, whereas the thermal adjustment timescale
in the convection zone, estimated using mixing-length theory,
is about 1 year.
We therefore assume $\fk(kD)^2 \gg 1$,
in which case
$G_1 \gg G_2,G_3$.
This same assumption was made implicitly in the models of
\citet{SpiegelZahn92} and \citet{GoughMcIntyre98},
in both of which the top of the tachocline was assumed to be isothermal.
Under this assumption we find that $G_1 \sim (kD)^3$,
and so (\ref{eq:w0estim2}) becomes
\begin{equation}
  w_{\rm t} \;\; \sim \;\; -\ii\bar{u}_{\rm cz}
\left(\dfrac{2\Ekappa}{n_{\rm rz}^2 kD^3}\right)\,.
  \label{eq:w0estim3}
\end{equation}
We note that $w_{\rm t}$ then depends only indirectly on the structure
of the interior magnetic field, via the tachocline thickness $D$.
The dependence of $w_{\rm t}$ on $D$
can be understood physically as a consequence of
thermal equilibrium and thermal-wind balance, as first
discussed by \citet{GoughMcIntyre98} \citep[see also][p~194]{McIntyre07}.
Since the vertical shear across the tachocline is of order $\bar{u}_{\rm cz}/D$,
thermal-wind balance implies that there must be a
temperature perturbation $\hat{T}$
of order $\bar{u}_{\rm cz}/(kD)$.
Such a perturbation cannot persist within the convection zone, where heat is
transported very efficiently by convective motions, but can persist within
the tachocline, where heat transport is less efficient.
In the tachocline, temperature perturbations diffuse at the rate $\Ekappa/D^2$,
and thermal equilibrium therefore requires that
$n^2_{\rm rz}w_{\rm t} \sim \Ekappa\hat{T}/D^2 \sim (\Ekappa/D^2)(\bar{u}_{\rm cz}/(kD))$,
from which (\ref{eq:w0estim3}) follows immediately.
Using order-of-magnitude estimates for the tachocline thickness, $D \simeq 0.01$,
and differential rotation, $\bar{u}_{\rm cz} \simeq 0.1$,
as well as $n_{\rm rz}^2/\Ekappa \simeq 8\times10^{13}$,
we find
from (\ref{eq:w0estim3})
that $w_{\rm t} \sim 10^{-9}$,
dimensionally $w_{\rm t} \sim 10^{-4}$ cm\,s$^{-1}$.
We note, however, that this result is rather sensitive to the tachocline
thickness $D$, which is not well constrained by observations.

Our model therefore
recovers \citeauthor{GoughMcIntyre98}'s scaling
for the meridional flow within the tachocline,
and goes further by
clarifying the physical
assumptions under which that scaling is derived,
and by identifying the necessary conditions for those
assumptions to hold.
Under more general conditions,
the meridional flow strength $w_{\rm t}$ is given by (\ref{eq:w0estim2}),
and (\ref{eq:w0estim3}) should be regarded instead as an approximate upper bound
on $w_{\rm t}$.
The more general formula (\ref{eq:w0estim2})
may be of relevance to
the interiors of solar-type stars that are more
weakly stratified, or that have thicker tachoclines.

Of course, we must be careful when applying the results of our idealized
model,
with its artificially confined magnetic field,
to the solar interior, or indeed to the interiors of other stars.
However,
since the formula for $w_{\rm t}$ given by (\ref{eq:w0estim3}) has no
explicit dependence on the structure of the background magnetic field $\bB_0$,
we argue that this result should hold for
more general, less artificial field configurations
than that considered here.
Moreover, the physical processes that give rise to (\ref{eq:w0estim3})
will be present in any self-consistent model of the solar interior,
and so (\ref{eq:w0estim3}) ought to be a robust result under solar-like conditions.

We now ask whether the meridional flows predicted by (\ref{eq:w0estim3})
would be of sufficient magnitude to confine an interior magnetic field against
outward diffusion.
In our idealized model, the interior field $\bB_0$ is confined by an
artificially imposed downwelling $\U$.
A first
approximation to the nonlinear magnetic confinement problem
can be obtained by setting $\U = |w_{\rm t}|$. Using (\ref{eq:U0def}),
we can then relate the tachopause thickness $\delta$
to the tachocline thickness $D$:
\begin{align}
\frac{\Eeta}{\delta} \;\; &\sim \;\; \bar{u}_{\rm cz}  \dfrac{2\Ekappa}{n_{\rm
    rz}^2 kD^3} \\
    \Rightarrow \;\; \delta \;\; &\sim \;\; \frac{\Eeta}{\Ekappa}
\frac{n_{\rm rz}^2 k}{2\bar{u}_{\rm cz} } D^3 \,.
\label{eq:deltaD}
\end{align}
This expression is equivalent to equation (7) in \citet{GoughMcIntyre98}.

We can now predict the strength of the magnetic field within the
tachopause, $B_{\rm t}$ say, using equation (\ref{eq:Bcrit}).
Here, our model differs significantly from that of \citeauthor{GoughMcIntyre98}.
They assumed that thermal-wind balance would hold within the tachopause,
as well as in the bulk of the tachocline,
and as a result the thickness of the tachopause in their model,
$\delta_{\rm GM}$ say,
was tied to the strength of the stable stratification.  They found that
\begin{equation}
  \delta_{\rm GM} = \left(
    \frac{2\Ekappa}{k^4n_{\rm rz}^2\Lambda_{\rm t}}
  \right)^{1/6}
  \label{eq:deltaGM}
\end{equation}
where $\Lambda_{\rm t} = \Lambda_{\rm loc}(z_0) = B_{\rm t}^2/(4\pi\eta\rho_0\Omega_\odot)$
is the Elsasser number within the tachopause.
In our model, on the other hand, the Lorentz force within the tachopause
overcomes thermal-wind balance, and we find that
\begin{equation}
  \delta =  \left(\frac{2}{k^2\Lambda_{\rm t}}\right)^{1/2}
  \label{eq:deltaUS}
\end{equation}
(see \S\ref{transition-layer_unstrat}).
Hence, whereas \citeauthor{GoughMcIntyre98} find a very strong dependence
of $B_{\rm t}$ on $D$, namely $B_{\rm t} \propto D^{-9}$ (see their
equation (8)), we find by combining (\ref{eq:deltaD})
and (\ref{eq:deltaUS}) that $B_{\rm t} \propto D^{-3}$, and more precisely,
\begin{equation}
\frac{B_{\rm t}^2}{4\pi \rho_0 R_\odot^2 \Omega_\odot^2} \sim \frac{
  8\bar{u}^2_{\rm cz}  }{ n_{\rm rz}^4 k^4 D^6 }
  \frac{E^2_\kappa}{\Eeta} \,.
  \label{eq:Btselect}
\end{equation}
Taking $k=2$, $n_{\rm rz}^2/\Ekappa \simeq 8\times10^{13}$,
and $\Eeta \simeq 3\times10^{-14}$, we find
\begin{align}
  \frac{B_{\rm t}^2}{4\pi \rho_0 R_\odot^2 \Omega_\odot^2}
  \;\; &\sim \;\; 3\times10^{-5} \left(\frac{\bar{u}_{\rm cz}}{0.1}\right)^2
    \left(\frac{D}{0.01}\right)^{-6} \\
  \Rightarrow \;\; B_{\rm t} \;\;
  &\sim \;\; 10^3 \left(\frac{\bar{u}_{\rm cz}}{0.1}\right)
    \left(\frac{D}{0.01}\right)^{-3}{\rm G}\,.
  \label{eq:Bt_estimate}
\end{align}
This estimate is rather higher than the $\sim 1$\,G field predicted by
\citeauthor{GoughMcIntyre98}.
We note that torsional Alfv\'enic oscillations of a $\sim1000$\,G field
could explain the 1.3 year oscillation detected in the tachocline's
angular velocity \citep{Gough00}.

\subsection{Guidance for the selection of parameters in numerical
models}

In recent years there have been several attempts to achieve magnetic
field confinement in self-consistent, nonlinear, global
numerical models of the solar interior.
Various approaches have been taken, including
axisymmetric steady-state calculations
\citep{GaraudGaraud08},
as well as two-dimensional and three-dimensional time-dependent simulations
\citep{GaraudRogers07,RogersMacGregor11,Strugarek-etal11}.
However, none of these numerical models has so far obtained
solutions that are satisfactorily close to solar observations,
and none has achieved magnetic confinement over an extended range
of latitudes.
These failures have
led some to conclude that the magnetic confinement picture
of \citeauthor{GoughMcIntyre98} is unworkable.
However, the analytical results
presented here suggest that magnetic confinement
by meridional flows
\emph{can} be achieved 
in the parameter regime appropriate to the solar tachocline.
We can also use our results
(a) to explain the lack of magnetic confinement in
existing numerical models,
and (b) to guide parameter
selection for future models.

As shown in \S\ref{sec:numexp_strat},
the amplitude of meridional flows within the radiative interior
decreases monotonically as the stratification is increased.
Moreover,
in a time-dependent model,
the timescale for the burrowing of the flows,
given by (\ref{eq:ES}), increases with stratification.
In the strongly stratified regime, with $\sigma \gtrsim 1$,
the burrowing of meridional flows is even slower than viscous diffusion.
But if $\sigma < 1$, as is the case in the tachocline, then the meridional
flows are approximately inviscid.
When modelling the solar interior, it is common practice to impose
a rotation rate and stratification profile that are close to solar,
in order to ensure a realistic separation between the dynamical timescales.
The viscous, thermal, and magnetic diffusivities ($\nu,\kappa,\eta$)
are then made as small as possible,
subject to computational constraints.
However, adopting this strategy does not guarantee that $\sigma < 1$.
Indeed, if the ratio of buoyancy and rotational frequencies $n_{\rm rz}$
is chosen to match the true tachocline value, then the condition
$\sigma < 1$ requires the Prandtl number $\nu/\kappa$ to be very small,
\begin{equation}
	\nu/\kappa < \frac{1}{n_{\rm rz}^2} \simeq 10^{-5}\,.
	\label{eq:sigma_condition}
\end{equation}
This condition is satisfied in the solar interior,
but is beyond the reach of current numerical simulations,
which instead typically have
a Prandtl number much closer to unity,
in order to reduce the numerical stiffness of the equations.
This places such simulations in the strongly stratified regime
in which (a) meridional flow velocity decays
exponentially with depth below the convection zone,
and (b) viscous stresses make a significant dynamical contribution,
even if the Ekman number is small, i.e.~even if $\Enu \ll 1$.

The
numerical
difficulty is, in fact, rather easily avoided by imposing a weaker
stratification, and thus allowing for a larger Prandtl number.
For example, if $n_{\rm rz} = 10$
then (\ref{eq:sigma_condition})
requires only that $\nu/\kappa < 10^{-2}$, which is readily
achievable numerically.  We then expect the meridional flow velocity
in the tachocline to scale as (\ref{eq:w0estim3}),
provided that the radiative--convective interface remains approximately
isothermal (see discussion below (\ref{eq:G1_G2})).

A further numerical difficultly is the need to resolve both the thin
tachocline and the thinner tachopause.  This difficultly can be
alleviated by allowing the tachocline to be somewhat thicker than is
observed in the Sun.\footnote{%
  The strength of the Sun's stratification increases rapidly with depth
  below the convection zone.  For models that use a solar-like vertical
  profile of $n_{\rm rz}$, care must be taken to ensure that the
  condition $\sigma < 1$ holds throughout the tachocline.
}
According to (\ref{eq:deltaD}),
to obtain a tachopause of thickness $\delta = 0.01$
and a tachocline of thickness $D = 0.1$, for example,
we require
\begin{equation}
  \frac{\Eeta}{\Ekappa}\frac{n_{\rm rz}^2 k}{2\bar{u}_{\rm cz}}
  \sim 10\,.
\end{equation}
If we choose $n_{\rm rz} = 10$, as suggested above,
as well as $\bar{u}_{\rm cz} \simeq 0.1$ and $k=2$,
then this requires a diffusivity ratio
$\eta/\kappa = \Eeta/\Ekappa \sim 10^{-2}$.
The strength of the interior magnetic field
must then be chosen in accordance with (\ref{eq:Btselect}), which requires
\begin{equation}
  \frac{B_{\rm t}^2}{4\pi \rho_0 R_\odot^2 \Omega_\odot^2}
  \sim 50\Ekappa\,.
\end{equation}
Finally, we must ensure that the Ekman length $\Enu^{1/2}$ is smaller
than $\delta = 0.01$.

All of the constraints just described
can be satisfied by choosing
$\Ekappa = 10^{-3}$, $\Eeta=10^{-5}$, and $\Enu = 10^{-6}$,
as well as $n_{\rm rz} = 10$
and $B_{\rm t}^2/(4\pi\rho_0) = 0.05 R_\odot^2 \Omega_\odot^2$.
These
parameter values
should be
numerically achievable;
in fact, the suggested
values
of $\Ekappa$, $\Eeta$, and $\Enu$
are very
similar to
those used by \citet{BrunZahn06}.

\section{Discussion and Conclusion}
\label{sec:disc}

Following the failure of several recent attempts to recreate the
\citeauthor{GoughMcIntyre98} tachocline scenario in global numerical
models,
our main goal in this work was to
identify in what parameter regime, if any, the results of
\citeauthor{GoughMcIntyre98} apply.
For this purpose we created a model of the tachocline
that is sufficiently simple to have analytical solutions yet, we
believe, incorporates
enough of the relevant dynamics to
yield quantitative predictions.

We have identified four distinct parameter
regimes that occur in our results
as the strength of the stratification is increased.
For solar parameter values, the results lie in either
the ``weakly stratified'' regime
or the ``moderately stratified'' regime
(see \S\ref{sec:weak_and_mod}).
In both of these regimes
the strength of the
meridional flow within the tachocline is determined
by a combination of thermal equilibrium and thermal-wind balance,
and follows the scaling predicted by \citeauthor{GoughMcIntyre98}.
With realistic tachocline parameters, the downwelling flow is of sufficient
strength to confine an interior magnetic field across a thin boundary layer,
and the thickness of the tachocline is related to the strength of the magnetic
field by (\ref{eq:Bt_estimate}).

By contrast,
all previous attempts to model the tachocline numerically
have been performed in the ``strongly'' stratified regime,
in which the burrowing of meridional flows is significantly
reduced by viscosity.
We believe
that this explains the lack of field confinement
in those models.
To remedy the problem, we suggest an alternative set of
numerically achievable parameters, which we predict will
yield results
that are
much more consistent with solar observations.

This project was initiated during the ISIMA 2010 summer program,
funded by the NSF CAREER grant 0847477, the France--Berkeley fund,
the Institute of Geophysics and Planetary Physics,
and the Center for the Origin, Dynamics and Evolution of Planets.
We thank them for their support.
P.G. and T.W. were also supported by NSF CAREER grant 0847477,
and J.M. was supported by NCAR and the Geophysical Turbulence Program.

\appendix

\section{Analytical derivation of the transition-layer solution}
\label{sec:app-analytic}

The magnetic transition layer
is described by equations (\ref{xmom2})--(\ref{divb2}),
except that (\ref{zmom2}) is replaced by (\ref{zmom3})
in cases with stratification.
After eliminating $\hat{p}$, $\huy$, and $\hby$,
these reduce to
\begin{eqnarray}
  2\frac{\dd\huz}{\dd z} &=& -k^2\Lambda\hbx\ee^{-z/\delta}
  \label{eq:reduce1} \\
  2\frac{\dd\hux}{\dd z} &=& -\ii kT_{\rm t} +
  \Lambda\left(
    \frac{1}{\delta^2}\hbz - \frac{\dd^2\hbz}{\dd z^2}
  \right)\ee^{-z/\delta} \label{eq:reduce2} \\
  0 &=& \ii k\hux\ee^{-z/\delta} + \frac{1}{\delta}\frac{\dd\hbx}{\dd z}
  + \frac{\dd^2\hbx}{\dd z^2} \\
  0 &=& \ii k\huz\ee^{-z/\delta} + \frac{1}{\delta}\frac{\dd\hbz}{\dd z}
  + \frac{\dd^2\hbz}{\dd z^2}\,.
  \label{eq:reduce4}
\end{eqnarray}
We now define a new variable $\zeta = \exp((z_0-z)/\delta)$,
with $z_0$ given by (\ref{eq:z0pred}), and so the domain
$-\infty < (z-z_0)/\delta < +\infty$ maps onto
$\infty > \zeta > 0$.
When written in terms of $\zeta$,
equations (\ref{eq:reduce1})--(\ref{eq:reduce4}) become
\begin{eqnarray}
  \frac{\dd\huz}{\dd\zeta} &=& k\sqrt{\frac{\Lambda}{2}}\,\hbx
  \label{eq:zeta1} \\
  \frac{\dd\hux}{\dd\zeta} &=& \frac{\ii k\delta T_{\rm t}}{2\zeta} +
  \frac{1}{k\delta^2}\sqrt{\frac{\Lambda}{2}}\left(
    \zeta^2\frac{\dd^2\hbz}{\dd\zeta^2} + \zeta\frac{\dd\hbz}{\dd\zeta} - \hbz
  \right) \label{eq:zeta2} \\
  \hux
  &=& \frac{\ii}{\delta}\sqrt{\frac{\Lambda}{2}}\,\zeta\frac{\dd^2\hbx}{\dd\zeta^2}
  \label{eq:zeta3} \\
  \huz
  &=& \frac{\ii}{\delta}\sqrt{\frac{\Lambda}{2}}\,\zeta\frac{\dd^2\hbz}{\dd\zeta^2}\,.
  \label{eq:zeta4}
\end{eqnarray}
After eliminating $\hbx$ and $\hbz$, we find
\begin{flalign}
	&& \ii k\delta\hux &= -\zeta\frac{\dd^3\huz}{\dd\zeta^3} & \label{eq:half1} \\
	& \mbox{and} & \ii k\delta\frac{\dd^2\hux}{\dd\zeta^2} &=
	\frac{k^2\delta^2T_{\rm t}}{2\zeta^2} + \zeta\frac{\dd\huz}{\dd\zeta} + 2\huz\,, &
	\label{eq:half2}
\end{flalign}
which we combine into a single equation for $\huz$,
\begin{equation}
	\left(\zeta\frac{\dd}{\dd\zeta}+2\right)\left(\frac{\dd^4}{\dd\zeta^4}+1\right)\huz
	= -\frac{k^2\delta^2T_{\rm t}}{2\zeta^2}\,.
	\label{eq:5th}
\end{equation}
This equation is equivalent to equation (\ref{eq:transition2}),
but expressed in terms of $\zeta$.
We are interested in the solution of (\ref{eq:5th}) that matches onto the flow
in the weakly magnetic region above,
and vanishes in the magnetically dominated region below.
These matching conditions can be written as
\begin{flalign}
  && \left.
  \begin{array}{rcl}
    \hux &\sim& u_{\rm t} + \frac{1}{2}\ii k\delta T_{\rm t}\ln\zeta \\
    \huz &\to& w_{\rm t} \\
    \hbx &\to& 0 \\
    \hbz &\to& 0
  \end{array}
  \right\} & \;\;\; \mbox{as} \;\;\; \zeta \to 0 &
  \label{eq:BC1} \\
  & \mbox{and} & \huz \to 0 & \;\;\; \mbox{as} \;\;\; \zeta \to \infty. &
  \label{eq:BC2}
\end{flalign}
Conditions (\ref{eq:BC1}) correspond to (\ref{eq:match}), plus the condition
that the magnetic field perturbation vanishes
above the transition layer.

After multiplying equation (\ref{eq:5th}) by $\zeta$,
and integrating once, we find
\begin{equation}
	\zeta^2\huz + \zeta^2\frac{\dd^4\huz}{\dd\zeta^4}
	= \mbox{const.} - \tfrac{1}{2}k^2\delta^2T_{\rm t}\ln\zeta\,. \label{eq:first1}
\end{equation}
Using (\ref{eq:half1}), we can
write this as
\begin{equation}
	\zeta^2\huz + \ii k\delta\left(\hux - \zeta\frac{\dd\hux}{\dd\zeta}\right)
	= \mbox{const.} - \tfrac{1}{2}k^2\delta^2T_{\rm t}\ln\zeta\,. \label{eq:first2}
\end{equation}
For compatibility with (\ref{eq:BC1}) the constant on the right-hand side
must equal $\ii k\delta u_{\rm t} + \tfrac{1}{2}k^2\delta^2T_{\rm t}$.
We can then use (\ref{eq:zeta2}) and (\ref{eq:zeta4}) to eliminate
$\dd\hux/\dd\zeta$ and $\huz$ from (\ref{eq:first2}), which leads to
\begin{equation}
	\hux - u_{\rm t} - \tfrac{1}{2}\ii k\delta T_{\rm t}\ln\zeta
	= \frac{1}{k\delta^2}\sqrt{\frac{\Lambda}{2}}\,
	\zeta^3\frac{\dd}{\dd\zeta}(\hbz/\zeta)\,.
	\label{eq:hux}
\end{equation}
The condition that $\hbz\to0$ as $\zeta\to0$ implies that the right-hand
side of (\ref{eq:hux}),
and hence also the left-hand side, must be $o(\zeta)$.
That is, both sides of (\ref{eq:hux}) must vanish faster than $\zeta$
as $\zeta\to0$.
Similarly, the condition that $\hbx\to0$ as $\zeta\to0$ implies that both
sides of (\ref{eq:zeta1}) are $o(1)$.
So we can use (\ref{eq:half1}) to express all four matching conditions in
(\ref{eq:BC1}) as
\begin{flalign}
  && \left.
  \begin{array}{rcl}
    \zeta\dfrac{\dd^3\huz}{\dd\zeta^3} &\sim& - \ii k\delta u_{\rm t}
    + \frac{1}{2}k^2\delta^2 T_{\rm t}\ln\zeta + o(\zeta) \\
    \huz &\sim& w_{\rm t} + o(\zeta)
  \end{array}
  \right\} & \;\;\; \mbox{as} \;\;\; \zeta \to 0. &
  \label{eq:BCshort}
\end{flalign}

Equation (\ref{eq:first1}), with the constant on the right-hand side
now identified as $\ii k\delta u_{\rm t} + \tfrac{1}{2}k^2\delta^2T_{\rm t}$,
becomes
\begin{equation}
	\huz + \frac{\dd^4\huz}{\dd\zeta^4} \;=\;
	\ii k\delta u_{\rm t}\,\frac{1}{\zeta^2} \;+\;
	\tfrac{1}{2}k^2\delta^2T_{\rm t}\,\frac{1-\ln\zeta}{\zeta^2}\,.
\end{equation}
We can write the general solution as
\begin{align}
  \huz &= \ii k\delta u_{\rm t}\,I_1(\zeta)
  + \tfrac{1}{2}k^2\delta^2T_{\rm t}\,I_2(\zeta)
  \nonumber \\ &\hspace{1cm}
  + c_1\exp\left(\tfrac{1+\ii}{\sqrt{2}}\zeta\right)
  + c_2\exp\left(\tfrac{-1-\ii}{\sqrt{2}}\zeta\right)
  + c_3\exp\left(\tfrac{1-\ii}{\sqrt{2}}\zeta\right)
  + c_4\exp\left(\tfrac{-1+\ii}{\sqrt{2}}\zeta\right)
  \label{eq:general}
\end{align}
where
$c_1$, ..., $c_4$ are arbitrary constants,
and
$I_1(\zeta)$ and $I_2(\zeta)$ are the integrals
\begin{align}
  I_1(\zeta) &= \int_0^\infty\!\frac{s\,\dd s}{\ee^{\zeta s}}\,\frac{1}{1+s^4}\,,
  \label{eq:particular1} \\
  I_2(\zeta) &= \int_0^\infty\!\frac{s\,\dd s}{\ee^{\zeta s}}\,\frac{\gamma+\ln s}{1+s^4}\,.
  \label{eq:particular2}
\end{align}
In equation (\ref{eq:particular2})
$\gamma$ is the Euler--Mascheroni constant,
\begin{equation}
	\gamma = -\int_0^\infty\ee^{-s}\ln s\,\dd s = 0.577...
\end{equation}

The values of the constants $c_1$, ..., $c_4$ are fixed by the
matching conditions (\ref{eq:BC2}) and (\ref{eq:BCshort}).
To determine their values we need to consider
the asymptotic behavior of $I_1$ and $I_2$.
It can be shown that
\begin{flalign}
  &&\left.
    \begin{array}{rcl}
      I_1
      &\sim& \frac{\pi}{4} - \frac{\pi}{2\sqrt{2}}\zeta
      \\
      I_2 &\sim&
      \gamma\frac{\pi}{4} - \frac{\pi}{2\sqrt{2}}(\gamma + \tfrac{\pi}{4})\zeta
      \\
      \zeta\dfrac{\dd^3I_1}{\dd\zeta^3} &\sim&
      - 1 + \frac{\pi}{2\sqrt{2}}\zeta
      \\
      \zeta\dfrac{\dd^3I_2}{\dd\zeta^3} &\sim& \ln{\zeta}
      + \frac{\pi}{2\sqrt{2}}(\gamma - \frac{\pi}{4})\zeta
    \end{array}
  \right\}
  \;\;\; &\mbox{as} \;\;\; \zeta \to 0&
  \label{eq:log} \\
  &\mbox{and}&\left.
    \begin{array}{rcl}
      I_1 &\sim& 1/\zeta^2 \\
      I_2 &\sim& (1 - \ln{\zeta})/\zeta^2
    \end{array}
  \right\}
  \;\;\; &\mbox{as} \;\;\; \zeta \to \infty\,.&
  \label{eq:infty}
\end{flalign}
Matching condition (\ref{eq:BC2}), together with (\ref{eq:infty}),
implies that $c_1 = c_3 = 0$.
The remaining matching conditions (\ref{eq:BCshort}), together with (\ref{eq:log}),
then imply that
\begin{flalign}
  && w_{\rm t} &= \ii\tfrac{\pi}{4}k\delta u_{\rm t}
     + \tfrac{\pi}{8}\gamma k^2\delta^2 T_{\rm t}
     + c_2 + c_4\,, &
  \label{eq:simul1} \\
  && 0 &= \ii\tfrac{\pi}{2\sqrt{2}}k\delta u_{\rm t}
    + \tfrac{\pi}{4\sqrt{2}}(\gamma + \tfrac{\pi}{4})k^2\delta^2 T_{\rm t}
    + c_2\ee^{\ii\pi/4} + c_4\ee^{-\ii\pi/4}\,, &
  \label{eq:simul2} \\
  & \mbox{and} & 0 &= \ii\tfrac{\pi}{2\sqrt{2}}k\delta u_{\rm t}
    + \tfrac{\pi}{4\sqrt{2}}(\gamma - \tfrac{\pi}{4})k^2\delta^2 T_{\rm t}
    + c_2\ee^{-\ii\pi/4} + c_4\ee^{\ii\pi/4}\,. &
  \label{eq:simul3}
\end{flalign}
Solving the three equations
(\ref{eq:simul1})--(\ref{eq:simul3})
fixes the values of $c_2$ and $c_4$, and also imposes a
condition on $w_{\rm t}$,
\begin{equation}
	w_{\rm t} = -\ii\tfrac{\pi}{4}k\delta u_{\rm t}
     - \tfrac{\pi}{8}\gamma k^2\delta^2 T_{\rm t}\,.
\label{eq:blsol-app}
\end{equation}
The solution for $\huz$ is
\begin{align}
	\huz &= \ii k\delta u_{\rm t}\left[\raisebox{0.5cm}{}
	  I_1(\zeta) - \tfrac{\pi}{2}\,{\rm Re}\left\{
	    \exp\left(-\tfrac{1+\ii}{\sqrt{2}}\zeta\right)
	  \right\}
  \right] \nonumber \\
  &\hspace{1cm}+
  \tfrac{1}{2}k^2\delta^2 T_{\rm t}\left[\raisebox{0.5cm}{}
	  I_2(\zeta) - \tfrac{\pi}{2}\,{\rm Re}\left\{
	    (\gamma-\ii\tfrac{\pi}{4})\exp\left(-\tfrac{1+\ii}{\sqrt{2}}\zeta\right)
	  \right\}
  \right]\,.
\end{align}
We can also calculate the change in the temperature gradient across the
transition layer.  This is given approximately by (\ref{eq:heating}),
and more precisely by
\begin{align}
	\left[\frac{\dd\hat{T}}{\dd z}\right]_{z=-\infty}^{z=+\infty}
	\;\; &= \;\;
	\frac{n_{\rm rz}^2}{\Ekappa}\left(
	  \int_{z=-\infty}^{z=+\infty}\huz\,\dd z
	  -	\int_{z=z_0}^{z=+\infty}w_{\rm t}\,\dd z
	\right) \\
	=& \;\; \frac{n_{\rm rz}^2\delta}{\Ekappa}\left(
	  \int_{0}^{\infty}\huz\,\frac{\dd\zeta}{\zeta}
	  -	\int_{0}^{1}w_{\rm t}\,\frac{\dd\zeta}{\zeta}
	\right) \\
	&= \;\; \frac{n_{\rm rz}^2\delta}{\Ekappa}\left(
	  \left(\tfrac{\pi}{4}\right)^3\tfrac{1}{2}k^2\delta^2 T_{\rm t}
	  - \gamma w_{\rm t}
	\right)\,.
	\label{eq:heating_better}
\end{align}

\section{The global solution, and the vertical flow velocity in the tachocline}
\label{sec:app-global}

As in the work of \citetalias{GaraudBodenheimer10},
we construct an approximate global solution of (\ref{xmom})--(\ref{divb}) by 
finding the general solution
in each region of the domain
and then matching these solutions across the boundaries.
In our case
the boundaries between the regions are at $z=h$ and $z=z_0$,
and are known \emph{a priori},
with $h$ given by the background stratification and $z_0$ given by
(\ref{eq:z0pred}).

\subsection{The general solutions in each region}

\subsubsection{Solution in the convection zone, $z \in [h,1]$}

In the convection zone the governing equations are well approximated by 
\begin{align}
-2\huy &= - \frac{\hux - u_{\rm cz}(z)}{\tau_c} \label{eq:cz-x} \\
2\hux &= -\ii k\hat{p} - \frac{\huy}{\tau_c} \label{eq:cz-y} \\
0 &= -\frac{\dd\hat{p}}{\dd z} + \hat{T} - \frac{\huz}{\tau_c} \label{eq:cz-z} \\
0 &= \ii k\huy + \frac{\dd\huz}{\dd z} \\
0 &= \frac{\dd^2\hat{T}}{\dd z^2} - k^2 \hat{T}\,.
\end{align}
The general solution for the temperature perturbation
can be written as
\begin{equation}
  \hat{T} = a \cosh k(z-1) + b \sinh k(z-1)
\end{equation}
where $a$ and $b$ are integration constants.
From the boundary condition $T=0$ at $z=1$ we deduce
immediately that $a=0$.
Combining the remaining equations yields
\begin{equation}
  \left(\frac{\dd^2}{\dd z^2} - \frac{1}{d^2}\right)\huz
  =
  \left[-\hat{T} + \frac{2\ii}{k}\frac{\dd u_{\rm cz}}{\dd z}\right]\frac{\tau_c}{d^2}
\end{equation}
where $d$ is the lengthscale defined in equation (\ref{eq:d}).
We write the general solution for $\huz$ as
\begin{equation}
  \huz =
  - \frac{2\ii\tau_c}{kd^2}\int_z^1\!\!\dd z'\,u_{\rm cz}(z')\cosh\left(\frac{z'-z}{d}\right)
  - \frac{b}{4\tau_c}\sinh k(z-1)
  + A\cosh\left(\frac{z-1}{d}\right) + B\sinh\left(\frac{z-1}{d}\right)
  \label{eq:cz-w}
\end{equation}
where $A$ and $B$ are two additional integration constants.
The boundary condition $\hat{w}=0$ at $z=1$ implies that $A=0$.
(Since we have neglected the viscous terms in (\ref{eq:cz-x})--(\ref{eq:cz-z})
we cannot impose the stress-free boundary condition at $z=1$.
Including the viscous terms would
lead to an Ekman-type boundary layer forming at $z=1$, but the
effect on the solution within the bulk of the convection zone
would be of order $\Enu \ll 1$.)

Finally, the pressure perturbation is found to be
\begin{align}
  \hat{p} &=
  \frac{2\ii}{k}u_{\rm cz} - \frac{d^2}{\tau_c}\frac{\dd\huz}{\dd z} \\
  &=
  - \frac{2\ii}{kd}\int_z^1\!\!\dd z'\,u_{\rm cz}(z')\sinh\left(\frac{z'-z}{d}\right)
  + \frac{bkd^2}{4\tau_c^2}\cosh k(z-1) - \frac{Bd}{\tau_c}\cosh\left(\frac{z-1}{d}\right)\,.
\end{align}

\subsubsection{Solution in the tachocline, $z \in [z_0,h]$}

Within the tachocline
we have
\begin{align}
-2\huy &= 0 \label{eq:no_pump} \\
2\hux &= -\ii k\hat{p} \\
0 &= -\frac{\dd\hat{p}}{\dd z} + \hat{T} \\
0 &= \ii k\huy +\frac{\dd\huz}{\dd z} \label{eq:mass-tc} \\
\frac{n_{\rm rz}^2}{\Ekappa}\huz &= \frac{\dd^2\hat{T}}{\dd z^2} - k^2 \hat{T}\,.
\end{align}
(\ref{eq:no_pump}) and (\ref{eq:mass-tc})
together imply that $\huz$ is a constant, $\huz = w_{\rm t}$ say.
The remaining equations can then be integrated, and the general solution
written as
\begin{eqnarray} 
\huz &=& w_{\rm t}  \nonumber \\
\hat{T} &=& \left(T_{\rm t} + \frac{n_{\rm rz}^2}{k^2\Ekappa} w_{\rm t}\right)\cosh k(z-z_0)
  + K \sinh k(z-z_0)
  - \frac{n_{\rm rz}^2}{k^2\Ekappa} w_{\rm t}  \nonumber \\
\hux &=& u_{\rm t} - \tfrac{1}{2}\ii \left(T_{\rm t} + \frac{n_{\rm rz}^2}{k^2\Ekappa} w_{\rm t}\right)\sinh k(z-z_0)
  - \tfrac{1}{2}\ii K(\cosh k(z-z_0) - 1)
  + \frac{\ii n_{\rm rz}^2w_{\rm t}}{2k \Ekappa}(z-z_0) \nonumber \\
\hat{p} &=& \frac{2\ii u_{\rm t}}{k}
  + \frac{1}{k}\left(T_{\rm t} + \frac{n_{\rm rz}^2}{k^2\Ekappa} w_{\rm t}\right)\sinh k(z-z_0)
  + \frac{K}{k}(\cosh k(z-z_0) - 1)
  - \frac{n_{\rm rz}^2w_{\rm t}}{k^2 \Ekappa}(z-z_0) \nonumber
\end{eqnarray}
where $u_{\rm t}$ and $T_{\rm t}$ are the values of $\hux$ and $\hat{T}$
at the bottom of the tachocline,
and $K$ is an additional integration constant.

\subsubsection{Solution in the magnetically-dominated region, $z<z_0$}

In this region the differential rotation and meridional flow both
vanish.
The temperature perturbation $\hat{T}$
therefore satisfies
\begin{align}
  0 &= \frac{\dd^2\hat{T}}{\dd z^2} - k^2\hat{T} \\
  \Rightarrow \hat{T} &= \alpha \cosh kz + \beta \sinh kz\,.
\end{align}
Since $\hat{T}=0$ at the lower boundary $z=0$, we must have $\alpha=0$.

\subsection{Matching conditions}
The values of the seven
integration constants $b$, $B$, $w_{\rm t}$, $u_{\rm t}$,
$T_{\rm t}$, $K$ and $\beta$ are now determined by applying matching conditions
across the interfaces $z=z_0$ and $z=h$.
At the radiative--convective interface, $z=h$,
we impose that $\huz$, $\hat{p}$,
$\hat{T}$, and $f(z)\frac{\dd\hat{T}}{\dd z}$ are all continuous.\footnote{
  By imposing that $\huz$ is continuous at $z=h$ we neglect any gyroscopic
  pumping within the overshoot region.  This is reasonable provided that the
  overshoot depth $\Delta$ is not too large.
}
These are the same continuity conditions imposed by \citetalias{GaraudBodenheimer10},
except that we use a more realistic thermal energy equation (\ref{fullenergy}),
and as a result we require the continuity of the heat flux, rather than
the temperature gradient.
Across the magnetic transition layer, at $z=z_0$, we impose continuity of $\hat{T}$
and the
relations (\ref{eq:blsol-app}) and (\ref{eq:heating_better})
derived from our transition-layer solution.

The matching conditions lead to the following seven relations
between the integration constants:
\begin{align}
  -\frac{2\ii\tau_c}{kd^2}\int_h^1\!\!\dd z'\,u_{\rm cz}(z')&\cosh\left(\frac{z'-h}{d}\right)
  + \frac{b}{4\tau_c}\sinh k(1-h) - B\sinh\left(\frac{1-h}{d}\right) \nonumber \\
  &= w_{\rm t} \label{eq:matching1} \\
  -\frac{2\ii}{kd}\int_h^1\!\!\dd z'\,u_{\rm cz}(z')&\sinh\left(\frac{z'-h}{d}\right)
  + \frac{bkd^2}{4\tau_c^2}\cosh k(1-h) - \frac{Bd}{\tau_c}\cosh\left(\frac{1-h}{d}\right)
  \nonumber \\
  &=
  \frac{2\ii u_{\rm t}}{k}
  + \frac{1}{k}\left(T_{\rm t} + \frac{n_{\rm rz}^2}{k^2\Ekappa} w_{\rm t}\right)\sinh kD
  + \frac{K}{k}(\cosh kD - 1)
  - \frac{n_{\rm rz}^2w_{\rm t}}{k^2\Ekappa}D \label{eq:matching2} \\
  -b\sinh k(1-h) &=
  \left(T_{\rm t} + \frac{n_{\rm rz}^2}{k^2\Ekappa} w_{\rm t}\right)\cosh kD
  + K\sinh kD - \frac{n_{\rm rz}^2w_{\rm t}}{k^2\Ekappa}
  \label{eq:matching3} \\
  bk\cosh k(1-h) &=
  \frac{k}{\fk}\left(T_{\rm t} + \frac{n_{\rm rz}^2}{k^2\Ekappa} w_{\rm t}\right)\sinh kD
    + \frac{k}{\fk}K\cosh kD
  \label{eq:matching4} \\
  T_{\rm t}
  &=
  \beta \sinh kz_0 \label{eq:matching5} \\
  Kk &= \beta k\cosh kz_0 +
  \frac{n_{\rm rz}^2\delta}{\Ekappa}\left(
	  \left(\tfrac{\pi}{4}\right)^3\tfrac{1}{2}k^2\delta^2 T_{\rm t}
	  - \gamma w_{\rm t}
	\right)  \label{eq:matching6} \\
  w_{\rm t} &= -\ii\tfrac{\pi}{4}k\delta u_{\rm t}
     - \tfrac{\pi}{8}\gamma k^2\delta^2 T_{\rm t}
  \label{eq:matching7}
\end{align}
where $D = h-z_0$ is the tachocline thickness.

We now seek an explicit expression for the value
of the vertical flow $w_{\rm t}$ within the tachocline.
We begin by eliminating $B$ between
(\ref{eq:matching1}) and (\ref{eq:matching2}),
which leads to
\begin{align}
  \frac{\ii kd}{\tau_c}w_{\rm t}&\coth\left(\frac{1-h}{d}\right)
  + \frac{\ii n_{\rm rz}^2w_{\rm t}}{k^2\Ekappa}kD
  - \ii\left(T_{\rm t} + \frac{n_{\rm rz}^2}{k^2\Ekappa} w_{\rm t}\right)\sinh kD
  - \ii K(\cosh kD - 1) \nonumber \\
  &=
  2\bar{u}_{\rm cz} - 2u_{\rm t}
  + \frac{\ii bkd}{4\tau_c^2}\left[
    \frac{\sinh k(1-h)}{\tanh (\frac{1-h}{d})} - kd\cosh k(1-h)
  \right] \label{eq:no-B}
\end{align}
where $\bar{u}_{\rm cz}$ is the weighted average
of the forcing in the convection zone defined by equation (\ref{eq:average}).
The remaining six integration constants,
including $w_{\rm t}$,
therefore depend on
$u_{\rm cz}(z)$ only through its average $\bar{u}_{\rm cz}$.
Hence the entire solution below the
convection zone is determined by $\bar{u}_{\rm cz}$.

Next, we combine (\ref{eq:matching5}) and (\ref{eq:matching6})
to eliminate $\beta$.  The result can then
be combined with (\ref{eq:matching3}) and (\ref{eq:matching4})
to express $b$, $T_{\rm t}$ and $K$ in terms
of $w_{\rm t}$.
The general result is rather complicated, so for simplicity we
describe only two limiting cases, which correspond to the
``weakly stratified'' and ``moderately stratified'' regimes
identified in \S\ref{sec:weak_and_mod}.

\subsubsection{The weakly stratified regime, $n_{\rm rz}^2/\Ekappa \ll 1/(k\delta^3)$}
In this regime we find that the
term in (\ref{eq:matching6}) involving $n_{\rm rz}$ becomes negligible.
Since this term arises from the change in the temperature gradient across the
transition layer,
this regime corresponds to the ``weakly stratified'' regime described in
\S\ref{sec:weak_and_mod}.  After neglecting this term, we find that
\begin{align}
  b \; &= \; \frac{n_{\rm rz}^2w_{\rm t}}{k^2\Ekappa}
    \frac{1 - \sech k(z_0+D) \cosh kz_0}
    {\fk\cosh k(1-h) \tanh k(z_0+D) + \sinh k(1-h)}
  \label{eq:b} \\
  T_{\rm t} \; &= \; -\frac{n_{\rm rz}^2w_{\rm t}}{k^2\Ekappa}
    \frac{\fk\tanh kz_0[1 - \sech kD]
      + \tanh k(1-h) \tanh kz_0 \tanh kD}
    {\fk[\tanh kz_0 + \tanh kD]
      + \tanh k(1-h)[\tanh kz_0 \tanh kD + 1]} \label{eq:T} \\
  K \; &= \; -\frac{n_{\rm rz}^2w_{\rm t}}{k^2\Ekappa}
    \frac{\fk[1 - \sech kD]
      + \tanh k(1-h) \tanh kD}
    {\fk[\tanh kz_0 + \tanh kD]
      + \tanh k(1-h)[\tanh kz_0 \tanh kD + 1]}
  \label{eq:K2}
\end{align}

\subsubsection{The moderately stratified regime, $n_{\rm rz}^2/\Ekappa \gg 1/(k\delta^3)$}
In this regime the transition-layer temperature $T_{\rm t}$ turns out to be smaller than the
value given by (\ref{eq:T}) by a factor $\Ekappa/(n_{\rm rz}^2k\delta^3) \ll 1$.
To good approximation, therefore, we can neglect $T_{\rm t}$ in each of the
matching conditions (\ref{eq:matching1})--(\ref{eq:matching7}), which
is equivalent to setting $z_0 = 0$ in (\ref{eq:b})--(\ref{eq:K2}).

After eliminating $u_{\rm t}$ between equations
(\ref{eq:no-B}) and (\ref{eq:matching7}), and applying
the formulae for $b$, $T_{\rm t}$ and $K$ just derived, we arrive
arrive at an explicit formula for $w_{\rm t}$,
\begin{equation}
  w_{\rm t} \;\; = \;\; \dfrac{-\ii\bar{u}_{\rm cz}}
    {\dfrac{kd}{2\tau_c}\coth\left(\dfrac{1-h}{d}\right) + \dfrac{4}{\pi k\delta}
      + \dfrac{n_{\rm rz}^2}{2k^2\Ekappa}[G_1 + G_2 - G_3]}
\end{equation}
where
\begin{align}
  G_1 &= kD
    - \frac{\fk[2 - 2\sech kD + \tanh kz_0 \tanh kD]
      + \tanh k(1-h) \tanh kD}
    {\fk[\tanh kz_0 + \tanh kD]
      + \tanh k(1-h)[\tanh kz_0 \tanh kD + 1]}
  \label{eq:G1-weak} \\
  G_2 &= \frac{kd}{k^2d^2-1}\left[
    kd - \frac{\tanh k(1-h)}{\tanh(\frac{1-h}{d})}
  \right]
    \frac{1 - \sech kh \cosh kz_0}
      {\fk\tanh kh + \tanh k(1-h)}
  \label{eq:G2-weak} \\
  G_3 &= \gamma k\delta\frac{\fk[1 - \sech kD]\tanh kz_0
      + \tanh k(1-h) \tanh kD \tanh kz_0}
    {\fk[\tanh kz_0 + \tanh kD]
      + \tanh k(1-h)[\tanh kz_0 \tanh kD + 1]}
  \label{eq:G3-weak}
\end{align}
in the weakly stratified regime, and
\begin{align}
  G_1 &= kD
    - \frac{\fk[2 - 2\sech kD]
      + \tanh k(1-h) \tanh kD}
    {\fk \tanh kD + \tanh k(1-h)}
  \label{eq:G1-mod} \\
  G_2 &= \frac{kd}{k^2d^2-1}\left[
    kd - \frac{\tanh k(1-h)}{\tanh(\frac{1-h}{d})}
  \right]
    \frac{1 - \sech kD}
      {\fk\tanh kD + \tanh k(1-h)}
  \label{eq:G2-mod} \\
  G_3 &\simeq 0
  \label{eq:G3-mod}
\end{align}
in the moderately stratified regime.


\begin{thebibliography}{41}
\providecommand{\natexlab}[1]{#1}

\bibitem[{{Basu} \& {Antia}(2003)}]{BasuAntia03}
{Basu}, S., \& {Antia}, H.~M. 2003, Astrophys. J., 585, 553

\bibitem[{{Bretherton} \& {Spiegel}(1968)}]{BrethertonSpiegel68}
{Bretherton}, F.~P., \& {Spiegel}, A.~E. 1968, Astrophys. J., 153, L77

\bibitem[{{Brown} et~al.(1989){Brown}, {Christensen-Dalsgaard}, {Dziembowski},
  {Goode}, {Gough}, \& {Morrow}}]{Brownal89}
{Brown}, T.~M., {Christensen-Dalsgaard}, J., {Dziembowski}, W.~A., {Goode}, P.,
  {Gough}, D.~O., \& {Morrow}, C.~A. 1989, Astrophys. J., 343, 526

\bibitem[{{Brun} \& {Zahn}(2006)}]{BrunZahn06}
{Brun}, A.~S., \& {Zahn}, J. 2006, A\&A, 457, 665

\bibitem[{{Charbonneau} et~al.(1999){Charbonneau}, {Christensen-Dalsgaard},
  {Henning}, {Larsen}, {Schou}, {Thompson}, \& {Tomczyk}}]{Charbonneaual99}
{Charbonneau}, P., {Christensen-Dalsgaard}, J., {Henning}, R., {Larsen}, R.~M.,
  {Schou}, J., {Thompson}, M.~J., \& {Tomczyk}, S. 1999, Astrophys. J., 527,
  445

\bibitem[{{Charbonnel} \& {Talon}(2005)}]{CharbonnelTalon05}
{Charbonnel}, C., \& {Talon}, S. 2005, Science, 309, 2189

\bibitem[{{Christensen-Dalsgaard} \& {Schou}(1988)}]{JCDSchou88}
{Christensen-Dalsgaard}, J., \& {Schou}, J. 1988, in Seismology of the Sun and
  Sun-Like Stars, ed. {E.~J.~Rolfe}, vol. 286 of \emph{ESA Special
  Publication}, 149--153

\bibitem[{{Elliott}(1997)}]{Elliott97}
{Elliott}, J.~R. 1997, A\&A, 327, 1222

\bibitem[{{Elliott} \& {Gough}(1999)}]{ElliottGough99}
{Elliott}, J.~R., \& {Gough}, D.~O. 1999, Astrophys. J., 516, 475

\bibitem[{{Ferraro}(1937)}]{Ferraro37}
{Ferraro}, V.~C.~A. 1937, MNRAS, 97, 458

\bibitem[{{Garaud}(2002)}]{Garaud02}
{Garaud}, P. 2002, MNRAS, 329, 1

\bibitem[{{Garaud} \& {Acevedo-Arreguin}(2009)}]{GaraudAA09}
{Garaud}, P., \& {Acevedo-Arreguin}, L. 2009, Astrophys. J., 704, 1

\bibitem[{{Garaud} \& {Bodenheimer}(2010)}]{GaraudBodenheimer10}
{Garaud}, P., \& {Bodenheimer}, P. 2010, Astrophys. J., 719, 313

\bibitem[{{Garaud} \& {Brummell}(2008)}]{GaraudBrummell08}
{Garaud}, P., \& {Brummell}, N.~H. 2008, Astrophys. J., 674, 498

\bibitem[{{Garaud} \& {Garaud}(2008)}]{GaraudGaraud08}
{Garaud}, P., \& {Garaud}, J. 2008, MNRAS, 391, 1239

\bibitem[{{Garaud} \& {Rogers}(2007)}]{GaraudRogers07}
{Garaud}, P., \& {Rogers}, T. 2007, in Unsolved Problems in Stellar Physics: A
  Conference in Honor of Douglas Gough, ed. {R.~J.~Stancliffe, J.~Dewi,
  G.~Houdek, R.~G.~Martin, \& C.~A.~Tout}, vol. 948 of \emph{American Institute
  of Physics Conference Series}, 237--248

\bibitem[{{Gilman} \& {Miesch}(2004)}]{GilmanMiesch04}
{Gilman}, P.~A., \& {Miesch}, M.~S. 2004, Astrophys. J., 611, 568

\bibitem[{{Gizon} et~al.(2010){Gizon}, {Birch}, \& {Spruit}}]{Gizonal10}
{Gizon}, L., {Birch}, A.~C., \& {Spruit}, H.~C. 2010, Annu. Rev. Astron.
  Astrophys., 48, 289

\bibitem[{{Gough}(2000)}]{Gough00}
{Gough}, D. 2000, Science, 287, 2434

\bibitem[{{Gough} \& {McIntyre}(1998)}]{GoughMcIntyre98}
{Gough}, D.~O., \& {McIntyre}, M.~E. 1998, Nature, 394, 755

\bibitem[{{Haber} et~al.(2002){Haber}, {Hindman}, {Toomre}, {Bogart}, {Larsen},
  \& {Hill}}]{Haber-etal02}
{Haber}, D.~A., {Hindman}, B.~W., {Toomre}, J., {Bogart}, R.~S., {Larsen},
  R.~M., \& {Hill}, F. 2002, Astrophys. J., 570, 855

\bibitem[{{Kippenhahn}(1963)}]{Kippenhahn63}
{Kippenhahn}, R. 1963, Astrophys. J., 137, 664

\bibitem[{{Kosovichev} et~al.(1997){Kosovichev}, {Schou}, {Scherrer}, {Bogart},
  {Bush}, {Hoeksema}, {Aloise}, {Bacon}, {Burnette}, {de Forest}, {Giles},
  {Leibrand}, {Nigam}, {Rubin}, {Scott}, {Williams}, {Basu},
  {Christensen-Dalsgaard}, {Dappen}, {Rhodes}, {Duvall}, {Howe}, {Thompson},
  {Gough}, {Sekii}, {Toomre}, {Tarbell}, {Title}, {Mathur}, {Morrison}, {Saba},
  {Wolfson}, {Zayer}, \& {Milford}}]{Kosovichev-etal97}
{Kosovichev}, A.~G., {Schou}, J., {Scherrer}, P.~H., {Bogart}, R.~S., {Bush},
  R.~I., {Hoeksema}, J.~T., {Aloise}, J., {Bacon}, L., {Burnette}, A., {de
  Forest}, C., {Giles}, P.~M., {Leibrand}, K., {Nigam}, R., {Rubin}, M.,
  {Scott}, K., {Williams}, S.~D., {Basu}, S., {Christensen-Dalsgaard}, J.,
  {Dappen}, W., {Rhodes}, E.~J., Jr., {Duvall}, T.~L., Jr., {Howe}, R.,
  {Thompson}, M.~J., {Gough}, D.~O., {Sekii}, T., {Toomre}, J., {Tarbell},
  T.~D., {Title}, A.~M., {Mathur}, D., {Morrison}, M., {Saba}, J.~L.~R.,
  {Wolfson}, C.~J., {Zayer}, I., \& {Milford}, P.~N. 1997, Sol.~Phys., 170, 43

\bibitem[{{Kumar} \& {Quataert}(1997)}]{KumarQuataert97}
{Kumar}, P., \& {Quataert}, E.~J. 1997, Astrophys. J., 475, L143+

\bibitem[{{MacGregor} \& {Charbonneau}(1999)}]{MacGregorCharbonneau99}
{MacGregor}, K.~B., \& {Charbonneau}, P. 1999, Astrophys. J., 519, 911

\bibitem[{{McIntyre}(1994)}]{McIntyre94}
{McIntyre}, M. 1994, in The Solar Engine and its Influence on Terrestrial
  Atmosphere and Climate, ed. {E.~Nesme-Ribes}, 293

\bibitem[{{McIntyre}(2007)}]{McIntyre07}
{McIntyre}, M.~E. 2007, in The Solar Tachocline, ed. {D.~W.~Hughes, R.~Rosner,
  \& N.~O.~Weiss}, 183

\bibitem[{{Mestel}(1953)}]{Mestel53}
{Mestel}, L. 1953, MNRAS, 113, 716

\bibitem[{{Mestel} \& {Weiss}(1987)}]{MestelWeiss87}
{Mestel}, L., \& {Weiss}, N.~O. 1987, MNRAS, 226, 123

\bibitem[{{Rogers} \& {MacGregor}(2011)}]{RogersMacGregor11}
{Rogers}, T.~M., \& {MacGregor}, K.~B. 2011, MNRAS, 410, 946

\bibitem[{{R\"udiger} \& {Kitchatinov}(1997)}]{RudigerKitchatinov97}
{R\"udiger}, G., \& {Kitchatinov}, L.~L. 1997, Astron. Nachr., 318, 273

\bibitem[{{R{\"u}diger} et~al.(2005){R{\"u}diger}, {Kitchatinov}, \&
  {Arlt}}]{Rudiger-etal05}
{R{\"u}diger}, G., {Kitchatinov}, L.~L., \& {Arlt}, R. 2005, A\&A, 444, L53

\bibitem[{{Schatzman}(1962)}]{Schatzman62}
{Schatzman}, E. 1962, Annales d'Astrophysique, 25, 18

\bibitem[{{Schou} et~al.(1998){Schou}, {Antia}, {Basu}, {Bogart}, {Bush},
  {Chitre}, {Christensen-Dalsgaard}, {di Mauro}, {Dziembowski}, {Eff-Darwich},
  {Gough}, {Haber}, {Hoeksema}, {Howe}, {Korzennik}, {Kosovichev}, {Larsen},
  {Pijpers}, {Scherrer}, {Sekii}, {Tarbell}, {Title}, {Thompson}, \&
  {Toomre}}]{Schou-etal98}
{Schou}, J., {Antia}, H.~M., {Basu}, S., {Bogart}, R.~S., {Bush}, R.~I.,
  {Chitre}, S.~M., {Christensen-Dalsgaard}, J., {di Mauro}, M.~P.,
  {Dziembowski}, W.~A., {Eff-Darwich}, A., {Gough}, D.~O., {Haber}, D.~A.,
  {Hoeksema}, J.~T., {Howe}, R., {Korzennik}, S.~G., {Kosovichev}, A.~G.,
  {Larsen}, R.~M., {Pijpers}, F.~P., {Scherrer}, P.~H., {Sekii}, T., {Tarbell},
  T.~D., {Title}, A.~M., {Thompson}, M.~J., \& {Toomre}, J. 1998, Astrophys.
  J., 505, 390

\bibitem[{{Spiegel} \& {Veronis}(1960)}]{SpiegelVeronis1960}
{Spiegel}, E.~A., \& {Veronis}, G. 1960, Astrophys. J., 131, 442

\bibitem[{{Spiegel} \& {Zahn}(1992)}]{SpiegelZahn92}
{Spiegel}, E.~A., \& {Zahn}, J. 1992, A\&A, 265, 106

\bibitem[{{Strugarek} et~al.(2011){Strugarek}, {Brun}, \&
  {Zahn}}]{Strugarek-etal11}
{Strugarek}, A., {Brun}, A.~S., \& {Zahn}, J. 2011, A\&A, submitted

\bibitem[{{Thompson} et~al.(1996){Thompson}, {Toomre}, {Anderson}, {Antia},
  {Berthomieu}, {Burtonclay}, {Chitre}, {Christensen-Dalsgaard}, {Corbard}, {De
  Rosa}, {Genovese}, {Gough}, {Haber}, {Harvey}, {Hill}, {Howe}, {Korzennik},
  {Kosovichev}, {Leibacher}, {Pijpers}, {Provost}, {Rhodes}, {Schou}, {Sekii},
  {Stark}, \& {Wilson}}]{Thompson-etal96}
{Thompson}, M.~J., {Toomre}, J., {Anderson}, E.~R., {Antia}, H.~M.,
  {Berthomieu}, G., {Burtonclay}, D., {Chitre}, S.~M., {Christensen-Dalsgaard},
  J., {Corbard}, T., {De Rosa}, M., {Genovese}, C.~R., {Gough}, D.~O., {Haber},
  D.~A., {Harvey}, J.~W., {Hill}, F., {Howe}, R., {Korzennik}, S.~G.,
  {Kosovichev}, A.~G., {Leibacher}, J.~W., {Pijpers}, F.~P., {Provost}, J.,
  {Rhodes}, E.~J., Jr., {Schou}, J., {Sekii}, T., {Stark}, P.~B., \& {Wilson},
  P.~R. 1996, Science, 272, 1300

\bibitem[{{Wood} \& {McIntyre}(2011)}]{WoodMcIntyre11}
{Wood}, T.~S., \& {McIntyre}, M.~E. 2011, J. Fluid Mech., in press
  (arXiv:1005.5482)

\bibitem[{{Zahn} et~al.(1997){Zahn}, {Talon}, \& {Matias}}]{Zahn-etal97}
{Zahn}, J., {Talon}, S., \& {Matias}, J. 1997, A\&A, 322, 320

\bibitem[{{Zhao} \& {Kosovichev}(2004)}]{ZhaoKosovichev04}
{Zhao}, J., \& {Kosovichev}, A.~G. 2004, Astrophys. J., 603, 776

\end{thebibliography}
\end{document}